\documentclass[fleqn,usenatbib]{mnras}

\usepackage{newtxtext,newtxmath}

\usepackage[T1]{fontenc}

\DeclareRobustCommand{\VAN}[3]{#2}
\let\VANthebibliography\thebibliography
\def\thebibliography{\DeclareRobustCommand{\VAN}[3]{##3}\VANthebibliography}

\usepackage{graphicx}	
\usepackage{amsmath}	

\defcitealias{Wu2015}{Paper I}
\defcitealias{Wu2017}{Paper II}
\defcitealias{Hsu2023}{Paper VIII}

\title[GMC Collisions. IX. Chemical Evolution]{GMC Collisions As Triggers of Star Formation. IX. Chemical Evolution}

\author[C.-J. Hsu et al.]{Chia-Jung Hsu,$^{1}$
Jonathan C. Tan,$^{1,2}$
Jonathan Holdship,$^{3,4}$
Duo, Xu,$^{2}$
Serena Viti,$^{3,4}$
Benjamin Wu,$^{5}$ 
\newauthor
and Brandt Gaches$^{1}$
\\
$^{1}$Department of Space, Earth \& Environment, Chalmers University of Technology, Gothenburg, Sweden \\
$^{2}$Dept. of Astronomy, University of Virginia, Charlottesville, Virginia 22904, USA \\
$^{3}$Leiden Observatory, Leiden University, PO Box 9513, 2300, RA Leiden, The Netherlands \\
$^{4}$Department of Physics and Astronomy, University College London, Gower Street, WC1E 6BT, London, UK \\
$^{5}$National Astronomical Observatory of Japan, National Institutes of Natural Sciences, 2-21-1 Osawa, Mitaka, Tokyo 181-8588, Japan
}

\date{Accepted XXX. Received YYY; in original form ZZZ}

\pubyear{2023}

\begin{document}
\label{firstpage}
\pagerange{\pageref{firstpage}--\pageref{lastpage}}
\maketitle

\begin{abstract}
Collisions between giant molecular clouds (GMCs) have been proposed as a mechanism to trigger massive star and star cluster formation. To investigate the astrochemical signatures of such collisions, we carry out 3D magnetohydrodynamics simulations of colliding and non-colliding clouds exposed to a variety of cosmic ray ionization rates (CRIRs), $\zeta$, following chemical evolution including gas and ice-phase components. 
At the GMC scale, carbon starts mostly in $\rm{C^+}$, but then transitions into C, CO, followed by ice-phase CO and $\rm{CH_3OH}$ as dense, cooler filaments, clumps and cores form from the clouds. The oxygen budget is dominated by O, CO and water ice. 
In dense regions, we explore the gas phase CO depletion factor, $f_D$, that measures the extent of its freeze-out onto dust grains, including dependence on CRIR and observables of mass surface density and temperature. We also identify dense clumps 
and analyze their physical and chemical properties, including after synthetic line emission modeling, investigating metrics used in studies of infrared dark clouds (IRDCs), especially abundances of CO, $\rm HCO^+$ and $\rm N_2H^+$. 
For the colliding case, we find clumps have typical densities of $n_{\rm H}\sim10^5\:{\rm{cm}}^{-3}$ and temperatures of $\sim20\:$K, while those in non-colliding GMCs are cooler. 
Depending on $\zeta$ and GMC dynamical history, we find CO depletion factors of up to $f_D\sim10$, and abundances of HCO$^+\sim 10^{-9}$ to $10^{-8}$ and $\rm{N_2H^+}\sim10^{-11}$ to $10^{-10}$. 
Comparison with observed IRDC clumps indicates a preference for low CRIRs ($\sim10^{-18}\:{\rm{s}}^{-1}$) and a more quiescent (non-colliding), cooler and evolved chemodynamical history. We discuss the general implications of our results and their caveats for interpretation of molecular cloud observations.

\end{abstract}




\section{Introduction}

Collisions between giant molecular clouds (GMCs) have been proposed as a mechanism to trigger formation of stars, including massive stars and star clusters, via compression of gas that is already dense and self-gravitating \citep[e.g.,][]{Scoville1986}. Shear-driven GMC collisions can also explain the dynamical Kennicutt-Schmidt relation if they are the dominant mode of star formation and such a mechanism is a natural way to link the kpc scales of galactic dynamics to the pc scales of clustered star-forming regions \citep{Tan2000}. Simulations of GMCs orbiting in galactic disks have found that this type of event happens frequently, i.e., every $\sim$ 10-20\% of a local orbital time \citep{Tasker2009, Dobbs2015, Li2018}. 

Observationally, several studies have found evidence for cloud collisions \citep[see the review of][]{Fukui2021}. A typical method involves identifying the ``bridge effect'' in the position-velocity diagram of CO emission lines \citep[e.g.,][]{Galvan-Madrid2010, Nakamura2012, Fukui2018, Dewangan2022}. Another approach is to search for evidence of large-scale shocks in and around GMCs, e.g., via SiO emission, which is expected to be sputtered from dust grains in shocked regions. Indeed, SiO has been found to be widespread in the vicinity of several infrared dark clouds (IRDCs) indicating the presence of large-scale shocks that may be the result of cloud collisions \citet{Jimenez-Serra2010, Cosentino2018}. 

Several simulation works have investigated the influence of collision velocity and magnetic field strength and orientation on the morphology \citep{Wu2017, Dobbs2021} and star formation rate \citep{Wu2020, Liow2020}.
To further understand the chemical abundances in these kinds of simulations, \citet{Bisbas2017} post-processed the results from \citet{Wu2017} with {\tt 3D-PDR} code and claimed that the C$^+$ fine-structure line can be useful to diagnose the event. In another approach, \citet{Priestley2021} post-processed their smooth particle hydrodynamics simulation results with {\tt UCLCHEM} astrochemical model and {\tt LIME} radiative transfer code. They claimed that NH$_3$ and HCN will have similiar bridging features as CO. However, these methods cannot trace the full chemodynamical evolution inside molecular clouds and can only post-process a limited spatial domain.

In this paper, we build upon the work of \citet{Wu2017} and extend the magnetohydrodynamics (MHD) simulation with an embedded chemical solver. This allows us to consider the influence of hydrodynamics on chemical evolution and how cloud collisions change the abundances of various species. The method is presented in \S\ref{sec:method}, results in \S\ref{sec:results}, and our conclusions in \S\ref{sec:conclusion}.

\section{Numerical Methods}
\label{sec:method}

\subsection{Astrochemical Model and Benchmark Tests}
\label{sec:astrochem-model}

To study the chemical evolution in the colliding clouds, we use a modified astrochemical network extracted from \textbf{\tt UCLCHEM (v1.3)} \citep{Holdship2017a}. The original chemical network in {\tt UCLCHEM} was extended from the gas-phase reactions from the {\tt UMIST} 2012 database \citep{Mcelroy2013} with freeze-out, several kinds of desorption reactions, including $\rm H_2$ desorption, photo-desorption, cosmic-ray desorption, and some approximated surface reactions. These freeze-out and desorption rates are determined by the model proposed in \citet{Rawlings1992} and \citet{Roberts2007}. The approximated surface reactions mostly assume that the species are immediately hydrogenated on a grain surface \citep[more details in][]{Holdship2017a}. In this study, we modify the photo-desorption model to align with the results of \citet{Walsh2012a} and also consider the contribution of cosmic-ray-induced thermal desorption proposed by \citet{Hasegawa1992}. For the binding energies used in the calculation of desorption rates, we used the same values used in the `Grid 2' model of \citet{Entekhabi2022}, i.e., values from {\tt UMIST} 2012, except CO binding energy is set to 1,100~K. We further limited the network to a selected set of 114 species, instead of the original 215, by omitting the S- and Cl-bearing species, as well as some complex molecules. The selected species are listed in Table~\ref{tbl:all-species} according to the number of atoms in each species. We have retained Si-bearing species since they are important for modeling of grain sputtering that produces SiO, which will be implemented in a follow-up paper to this work. We note that the temperature ranges of gas-phase reactions are not considered in the standard {\tt UCLCHEM} model. This is a minor effect for the cold gas ($< 30\:$K) that we are mostly interested in. However, we still considered them in this study as the temperature can reach quite warm values in the simulation, e.g., in shocks or in the external layers around the GMCs.

As is generally the case for astrochemical models, the model we are using requires the inputs of density ($n_{\rm H}$), temperature ($T$), visual extinction ($A_V$), UV radiation field ($F_{\rm UV}$) and cosmic-ray ionization rate (CRIR, $\zeta$). We note that the $\rm H_2$ and $\rm CO$ column densities used in the photo-dissociation reactions are directly derived from $A_V$ by $\Sigma_{\rm H2} = 0.5 \times 1.59 \times 10^{21} \times A_V$ and $\Sigma_{\rm CO} = 10^{-5} \times \Sigma_{\rm H2}$, which are different from the standard {\tt UCLCHEM} methods, but the same as in \citet{Walsh2015}. 

\begin{figure*}
    \centering
    \includegraphics[width=\linewidth]{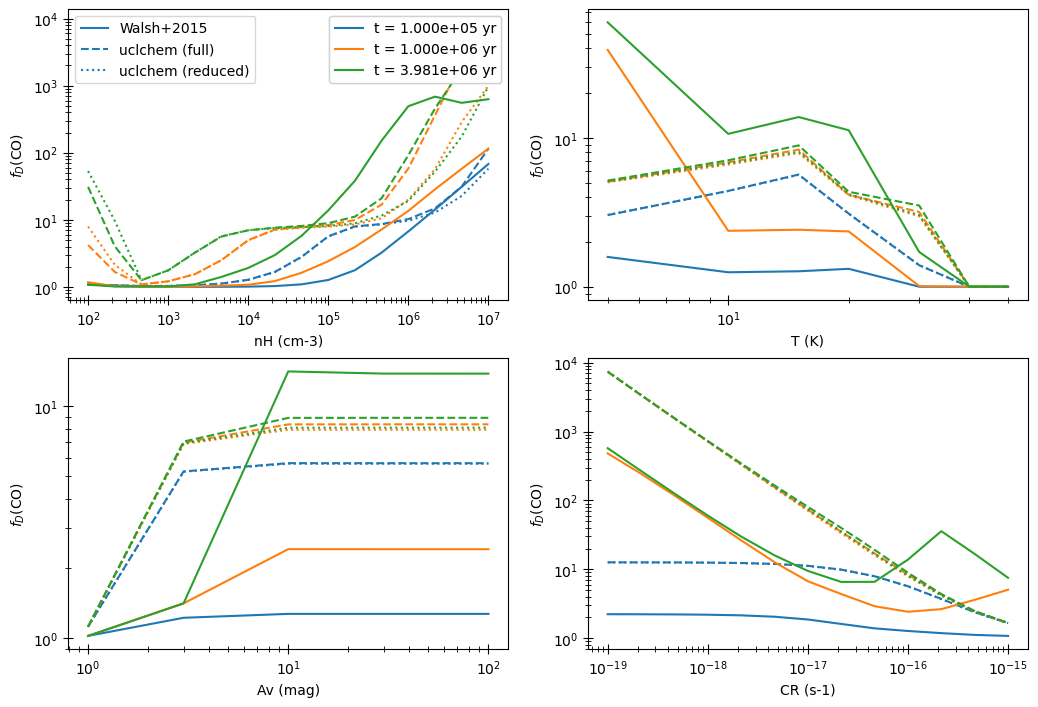}
    \caption{CO depletion factor benchmark test for a grid of models. Solid lines show the results of the `Grid 2' model of \citet{Entekhabi2022}. Dashed lines and dotted lines show results from the full {\tt UCLCHEM} network and the reduced {\tt UCLCHEM} network, respectively. The cosmic-ray desorption efficiency of these UCLCHEM models is set to $\phi = 10^5$. The different colors indicate the time evolution from $\sim 0.1$~Myr to $\sim 4$~Myr. {\it (a) Top left:} CO depletion factor versus density. {\it (b) Top right:} CO depletion factor versus temperature. {\it (c) Bottom left:} CO depletion factor versus visual extinction. {\it (d) Bottom right:} CO depletion factor versus cosmic-ray ionization rate.}
    \label{fig:benchmark_codep_phi1e5}
\end{figure*}

To validate the reduced modified network, we benchmarked against the original full chemical network and the `Grid 2' model of \citet{Entekhabi2022}. Figure~\ref{fig:benchmark_codep_phi1e5} shows the comparison of CO depletion factor $f_D({\rm CO})$ among the models up to around 4~Myr, which is the target time of our cloud collision simulation (see below). We sample: 16 values of density $n_{\rm H}$ equally in logarithmic space from 100 to $10^7$ ${\rm cm^-3}$; 7 values of temperature ($T$) of 5, 10, 15, 20, 30, 40 and 50~K; 5 values of visual extinction ($A_V$) of 1, 3, 10, 30, 100~mag; and 13 values of cosmic ray ionization rate $\zeta$ equally in logarithmic space from $10^{-19}$ to $10^{-15}\:\rm s^{-1}$. The UV radiation field ($F_{\rm UV}$) is fixed to 4 Habing field units, which is the value adopted by \citet{Wu2015} and \citet{Wu2017} for the PDR-based heating and cooling functions and is expected to be an appropriate value for GMCs and IRDCs in the inner Galaxy, e.g., at typical galactocentric radii of $\sim 4\:$kpc. The fiducial values of each parameter are $n_{\rm H} = 10^5\:{\rm cm^{-3}}$, $T=15\:{\rm K}$, $A_V = 30.0$ mag and $\zeta = 10^{-16}\:{\rm s^{-1}}$.

\begin{figure*}
    \centering
    \includegraphics[width=\linewidth]{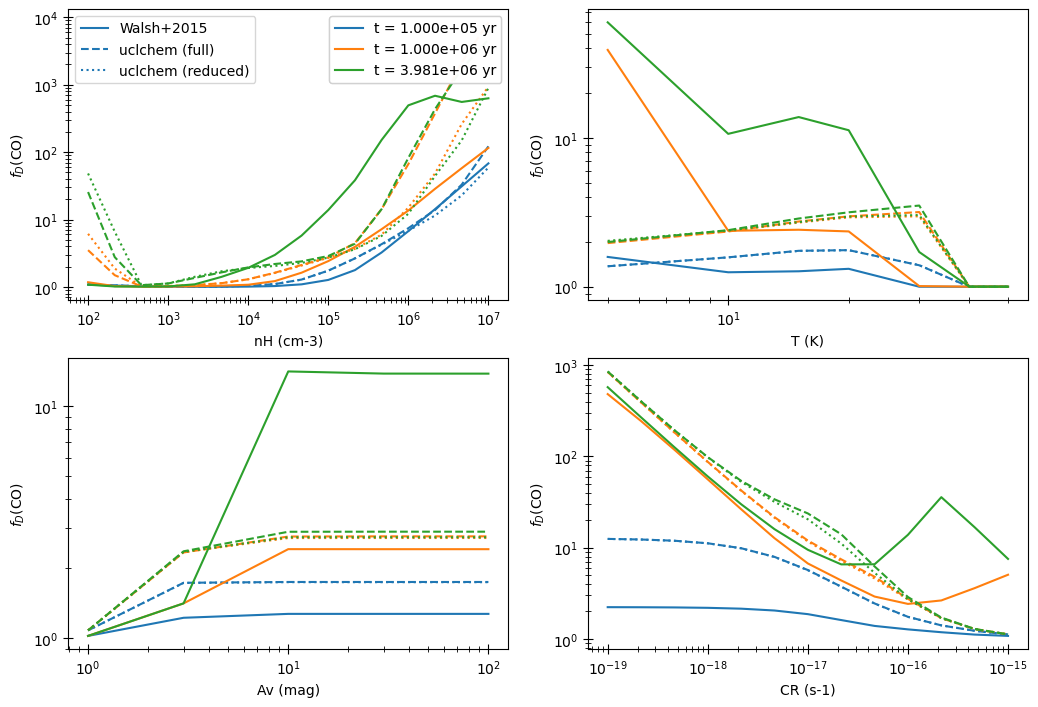}
    \caption{As Figure~\ref{fig:benchmark_codep_phi1e5}, but with cosmic-ray desorption efficiency $\phi = 10^8$ and maximum cosmic-ray desorption energy $E_{\rm t,CR} = 2000\:K$ (see text).}
    \label{fig:benchmark_codep}
\end{figure*}

Figure~\ref{fig:benchmark_codep_phi1e5} shows that the reduced network has a reasonable agreement with the full network. However, we can tell that the CO depletion factor from either the full or reduced UCLCHEM network has a systematically higher value than the `Grid 2' model in the medium density range and lower CRIR range, especially after 1~Myr. To obtain a better agreement, we change the cosmic-ray desorption efficiency ($\phi$) from $10^5$ to $10^8$, and change the maximum cosmic-ray desorption energy to $E_{\rm t,CR} = 2000\:K$. The details of these two parameters are described by \citet{Roberts2007}. The new benchmark results are shown in Figure~\ref{fig:benchmark_codep}. After we made this modification, we see that the difference of the CO depletion factor is usually within a factor of three by 1~Myr. After 1~Myr, the denser regions can show a higher difference, reaching up to a factor of 10. The phenomenon also appears in the original model without changing the cosmic-ray desorption efficiency. We note that in our simulations, the dense regions ($n_{\rm H} > 10^4\:(\rm cm^{-3})$ do not exist for longer than 1~Myr (see Section~\ref{sec:results}), so we expect their formation and early evolution will be reasonably well modeled by our adopted chemical network. 

\begin{table*}
    \centering
    \caption{Table of included species in the reduced astrochemical network.}
    \label{tbl:all-species}
    \begin{tabular}{lllllllllllll}
    \#atoms | & \multicolumn{9}{c|}{gas-phase} | & \multicolumn{3}{c|}{ice-phase} \\
    \hline
    1  & Mg      & Mg$^+$  & Si$^+$  & Si      & N$^+$   & O$^+$   & C$^+$   & He$^+$  & He     &  \#Mg    \\
       & N       & C       & H$^+$   & O       & e-      & H  & & & & & & \\
    2  & HeH$^+$ & SiC$^+$ & SiC     & SiH$^+$ & SiH     & SiO$^+$ & SiO     & CN$^+$  & N2$^+$ &  \#SiO  & \#CO   & \#NO   \\
       & CO$^+$  & H2$^+$  & NH$^+$  & O2$^+$  & NO$^+$  & N2      & CH$^+$  & NH      & OH$^+$ &  \#O2   & \#SiC  & \#N2   \\
       & CN      & CH      & NO      & OH      & O2      & CO      & H2      &         &        &        & & \\
    3  & HOC$^+$ & SiC2$^+$& SiC2    & O2H     & NO2     & SiH2$^+$& OCN     & SiH2   & \#HNC   & \#NO2  & \#O2H  \\
       & SiOH$^+$& HNO     & N2H$^+$ & O2H$^+$ & HNC     & HNO$^+$ & CO2     & HCN$^+$ & NH2$^+$& \#SiC2  & \#CO2  & \#HNO  \\
       & NH2     & CH2$^+$ & H2O$^+$ & HCN     & CH2     & HCO     & H3$^+$  & HCO$^+$ & H2O    & \#HCN   & \#H2O  & \\
    4  & SiC3$^+$& H2CN    & H2NO$^+$& H2SiO   & HNCO    & SiC3    & SiH3$^+$& SiH3    & HCO2+  & \#H2CN  & \#HNCO & \#SiC3 \\   & HCNH$^+$& NH3     & CH3     & CH3$^+$ & NH3$^+$ & H3O$^+$ & H2CO$^+$& H2CO    &        & \#H2SiO & \#H2CO & \#NH3  \\
    5  & SiH4$^+$& SiH4    & CH4$^+$ & H3CO$^+$& CH4     &         &         &         &        & \#CH4   & \#SiH4 & \\
    6  & SiH5$^+$& CH3OH   &         &         &         &         &         &         &        & \#CH3OH &        & \\
    \end{tabular}
    \bigskip
    \\
    \raggedright
\end{table*}

\subsection{Chemodynamical Simulation}

\subsubsection{Numerical Codes}

The simulations presented in this paper are based on the models introduced in \citet[][Paper II]{Wu2017} and the updated heating/cooling functions described in \citet[][Paper IV]{Christie2017}, while the codes are upgraded. We use the adaptive mesh refinement (AMR) code {\tt Enzo} v2.6 \citep{Brummel-Smith2019} to carry out the ideal magnetohydrodynamics (MHD) simulations. The ideal MHD equations are solved by MUSCL-Dedner method \citep{Dedner2002, Wang2008, Wang2009} with the HLLD Riemann solver \citep{Bryan2014}. Accompanying with the upgrade of {\tt Enzo}, the astrochemical library {\tt grackle} \citep{Smith2017} used to solve the heating and cooling rates in the simulations is also upgraded to v3.2.

To further study the chemical evolution in the simulations, we couple the reduced astrochemical model presented in Section~\ref{sec:astrochem-model} with {\tt Enzo} by using {\tt Naunet} (Hsu et al., in prep.). The ISM chemical model is solved in each cell of the simulation and the advection is handled by {\tt Enzo}. As the chemistry costs a lot of computational resources, but evolves more slowly than the hydrodynamics, we only solve the chemistry every 64 hydrodynamic timesteps and skip the substeps in the refined grids. In addition, to ensure elemental and charge conservation, we renormalize the species abundance every step with the method proposed by \citet{Grassi2017}. As we have uniform elemental abundances in the whole domain in the beginning, we simply apply this method to all the cells.

\subsubsection{Initial Conditions}

On the magnetohydrodynamics side, we follow the same settings as the fiducial case of \citetalias{Wu2017}. However, as clarified by \citet[][Paper VIII]{Hsu2023}, the initial density of the two uniform clouds is $n_{\rm H} = 83\:({\rm cm}^{-3})$ and the initial density of ambient gas is $n_{\rm H} = 8.3\:({\rm cm}^{-3})$. Each cloud has a radius $R_{\rm GMC} = 20$ pc. We note that the mass of each cloud is $9.4 \times 10^4 M_{\odot}$. The two clouds are initialized in a cubic domain with side length of 128 pc.  Each of them is initialized with temperature $T=15$~K and the ambient gas, whose density is 10 times lower, is initialized with 10 times higher temperature $T=150$~K to keep pressure balance. A solenoidal turbulent velocity field with power spectrum following $v_k^2 \propto k^{-4} (2 \leq k \leq 20$) is initialized in each cloud. A uniform, relatively weak magnetic field $B = 10\mu {\rm G}$ is applied in the whole domain in a direction to the 60$^\circ$ with respect to the collision axis. For the colliding case, the two clouds move towards each other with a velocity of 5 km/s along the collision axis, i.e., a relative velocity of 10~km/s. Note, the clouds are offset from each other along this axis by $0.5R_{\rm GMC}$. For the non-colliding case, their velocities are set to zero.

On the chemistry side, to be consistent with the heating/cooling table, which was generated from {\tt PyPDR} code \citet[][Paper I]{Wu2015}, we have a homogeneous background radiation field $F_{\rm UV} = 4 {G_0}$, i.e., 4 Habing fields. Similarly, the cosmic-ray ionization rate (CRIR, $\zeta$) is set to $10^{-16}\:{\rm s}^{-1}$ in the fiducial case. Visual extinction ($A_V$) is another required parameter of the astrochemical model, and we follow the $n_{\rm H}-A_V$ relation proposed in \citetalias{Wu2015}. The initial abundances are listed in Table~\ref{tab:initial-abund}, which are the same values as used in the study of \citet{Entekhabi2022}.

We note that in general our adopted heating/cooling rates will not be fully consistent with the modeled, evolving abundances of various chemical species. Future work is need to achieve such dynamic coupling of heating/cooling rates to local current abundances. Similarly, we also note that when we study chemical evolution under different CRIRs, the heating/cooling rates are still those tabulated from the fiducial value, i.e., there is always the same physical model being run, independent of the chemistry. While this is a simplification of reality, it does have the advantage of allowing us explore differences that are due purely to the astrochemical modeling aspects of the CRIR.

\begin{table}
    \centering
    \begin{tabular}{lc}
    \hline
        Species & Abundance ($n_{\rm species}/n_{\rm H}$) \\
    \hline
        {$\rm H_2$} & 1.0 \\
        {\rm H} & $5.0 \times 10^{-5}$ \\ 
        {\rm He} & $9.75 \times 10^{-2}$ \\ 
        {\rm N} & $7.5 \times 10^{-5}$ \\ 
        {\rm O} & $1.8 \times 10^{-4}$ \\ 
        {\rm CO} & $1.4 \times 10^{-4}$ \\ 
        {\rm Mg} & $7.0 \times 10^{-9}$ \\ 
        {\rm Si} & $8.0 \times 10^{-9}$ \\ 
    \hline
    \end{tabular}
    \caption{The initial abundances of species. Except for C that is initialized in CO and H in H$_2$, other species start from atomic form. We use the same values as \citet{Entekhabi2022} if the species exists in both models. However, Mg only exists in the {\tt UCLCHEM} model.}
    \label{tab:initial-abund}
\end{table}

The simulation domain is resolved by $128^3$ cells on a base grid. Then cells are able to be refined further with the condition to resolve the local Jeans length by 8 cells. We set the maximum number of levels of refinement to be four. This helps us achieve another $2^4$ times higher resolution, i.e., an equivalent resolution of 0.0625~pc. 

\subsection{Line emission\label{sec:synthetic}}

To make predictions that are most relevant to observational studies, we utilize the {\tt RADMC-3D} radiative transfer code \citep{2012ascl.soft02015D} to simulate the intensity of various emission lines from the modeled chemical species. Due to computational memory constraints, for this aspect, we focus on a 12~pc$^3$ box extracted from defined ``high density'' regions (see below). 
With density, temperature, velocity and molecular abundances of CO, HCO$^+$ and N$_2$H$^+$ output from the simulations, we model the emission of C$^{18}$O(1-0), H$^{13}$CO$^+$(1-0) and N$_2$H$^+$(1-0) rotational transitions, chosen to match the main species analyzed in the observational study of \citet{Entekhabi2022}. Note, for this calculation, we assume that the dust temperature is equal to the gas temperature and that H$_2$ is the collisional partner for all targeted molecules. The isotope ratios of $^{12}$C$/^{13}$C and $^{16}$O$/^{18}$O are assumed to be 50 \citep{Zeng2017} and 320 \citep{Hezareh2008}.



\section{Results}
\label{sec:results}

Similar to \citetalias{Wu2017}, our analysis is done in a rotated frame ($x', y', z'$) whose $z'$-axis is in the (15$^\circ$, 15$^\circ$) direction of the original ($r, \theta, \phi$) coordinate. This is done to avoid emphasis in the images of the contact boundary between the two uniform density ambient media, which are also moving with the GMCs.

\subsection{Density and Temperature}

\begin{figure}
    \centering
    \includegraphics[width=\linewidth]{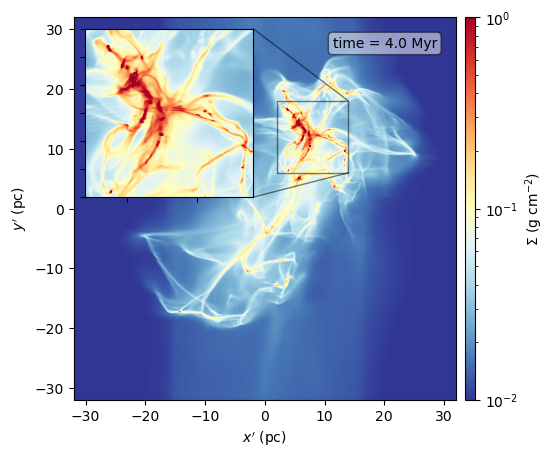}
    \includegraphics[width=\linewidth]{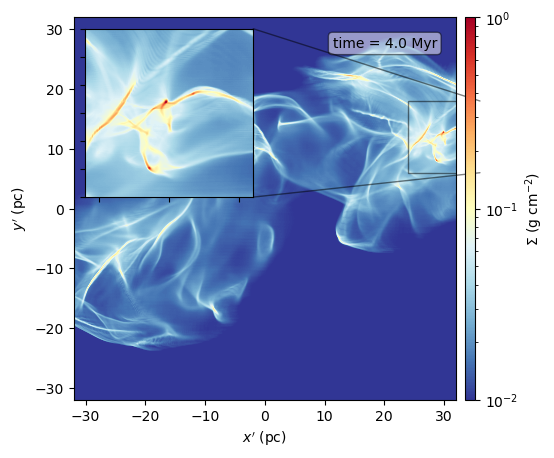}
    \caption{Mass surface density map and a zoom-in region of the colliding and non-colliding clouds at 4~Myr.}
    \label{fig:im-dens-zoom}
\end{figure}

In Figure~\ref{fig:im-dens-zoom}, we show the mass surface density maps of the colliding and non-colliding clouds at 4~Myr. In the former, the zoom-in area shows the center of the collision, which contains most of dense structures. Later discussions will focus on this dense region. A dense region is also shown as a zoom-in for the non-colliding case.

\begin{figure*}
    \centering
    \includegraphics[width=\linewidth]{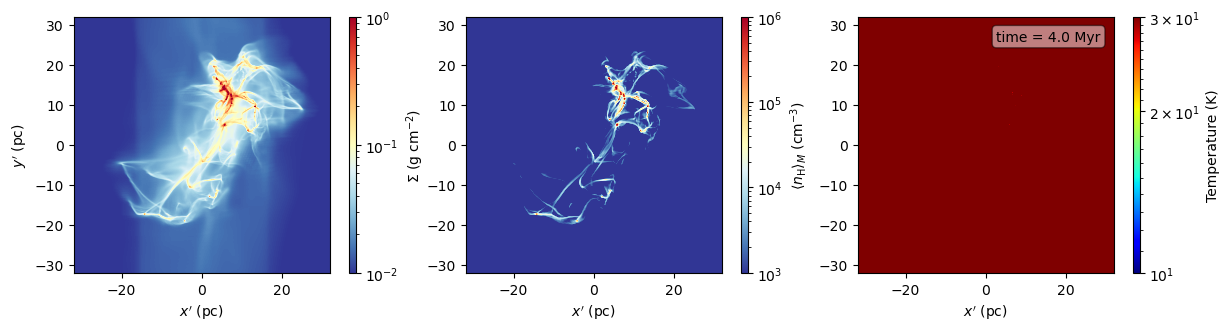}
    \includegraphics[width=\linewidth]{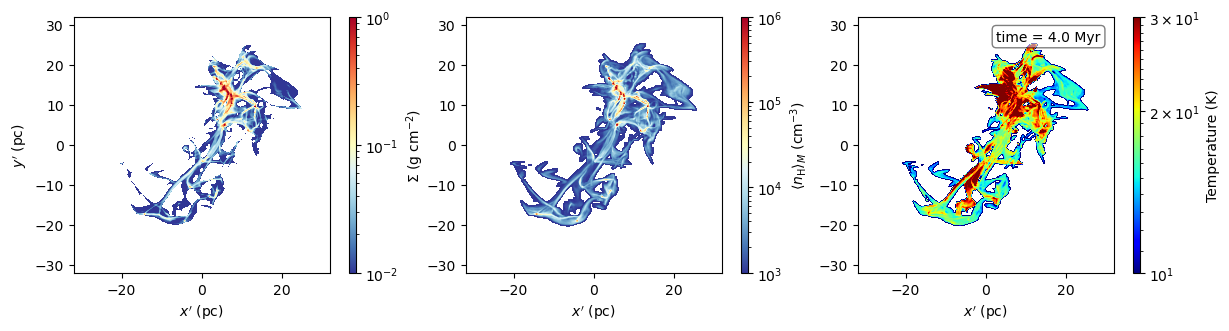}
    \includegraphics[width=\linewidth]{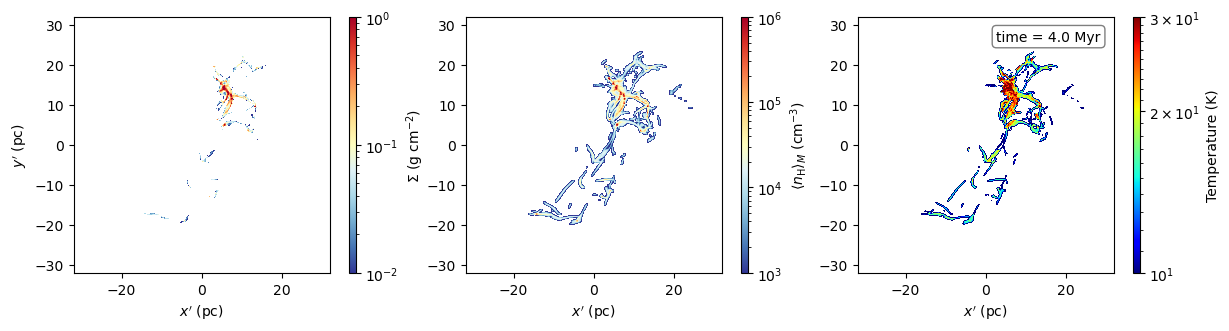}
    \caption{The maps of surface density (left column), mass-weighted density (middle column) and mass-weighted temperature (right column) of the colliding case at 4~Myr. From top to bottom, each row shows the maps under different density thresholds ($n_{\rm H} = 0$, $10^3$, and $10^4\:{\rm cm^{-3}}$).}
    \label{fig:im-dens-temp}
\end{figure*}

Figure~\ref{fig:im-dens-temp} shows projected maps (through the whole domain) of the colliding case, i.e., mass surface density, mass-weighted density and mass-weighted temperature. Different rows in the figure show the effect of using different density thresholds with which to select cells for inclusion in the analysis. From top to bottom, we set the density threshold to 0 (i.e., no density threshold), $n_{\rm H} = 10^3\:{\rm cm^{-3}}$, and $n_{\rm H} = 10^4\:{\rm cm^{-3}}$. Although the density outside of the clouds is 10 times smaller than in the initial clouds, the large domain still causes the contributions from the surrounding ambient medium to be non-negligible. In the mass surface density map, we can see that the merged cloud usually has $\Sigma>0.01\:{\rm g\:cm^{-2}}$ if the ambient gas is included (first row). However, we see that the dense structures only occupy a limited region from the mass-weighted density map. The ambient gas also causes the average temperature to be typically $>30$~K, except in some small localised regions embedded in the dense structures. However, when we apply the density threshold of $n_{\rm H} = 10^3\:{\rm cm^{-3}}$, we can see a significant drop in the mass surface density and temperature maps. After further increasing the density threshold to $n_{\rm H} = 10^4\:{\rm cm^{-3}}$, we have filtered out most of the ambient gas and only the localized dense structures remain. We consider this case with highest density threshold applied to be most relevant for seeing the localised conditions of prestellar clumps and cores. However, we note that when we select and analyze a high density region, below, that has a more limited spatial extent of 12~pc on a side, then we will revisit the choice and need for a density threshold.

\begin{figure*}
    \centering
    \includegraphics[width=\linewidth]{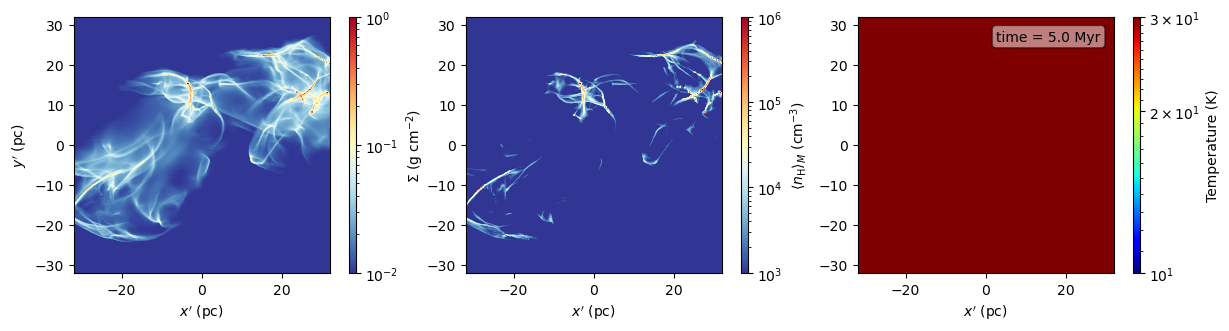}
    \includegraphics[width=\linewidth]{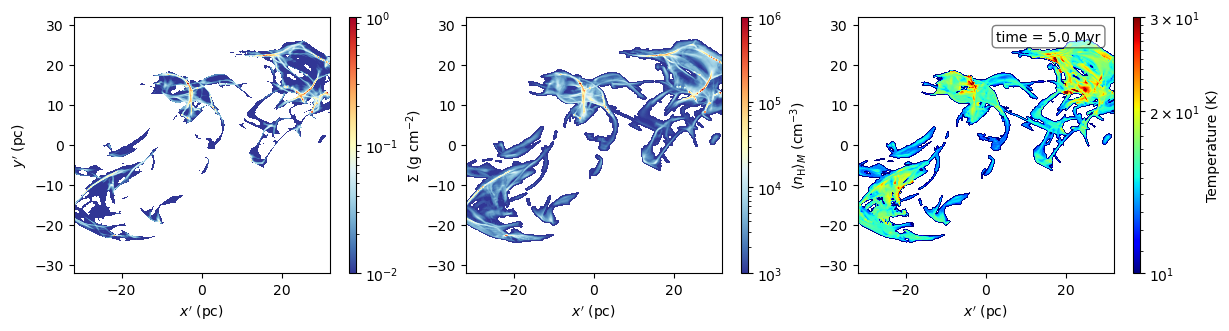}
    \includegraphics[width=\linewidth]{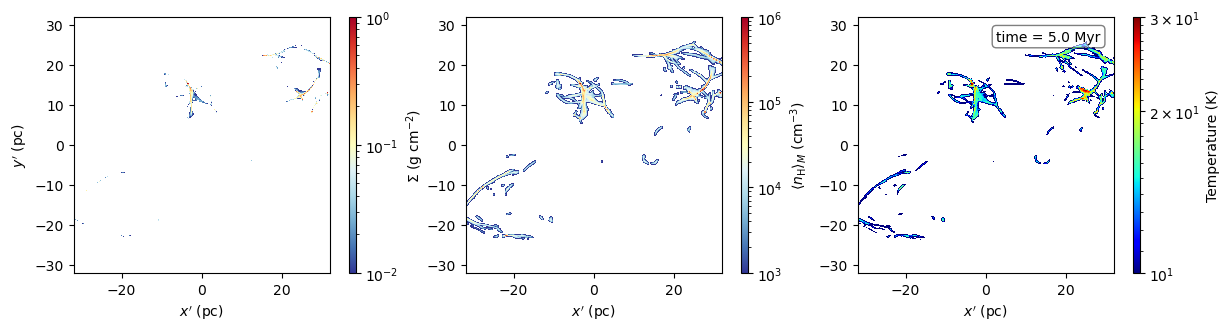}
    \caption{As Figure~\ref{fig:im-dens-temp}, but for the non-colliding case at 5~Myr.}
    \label{fig:im-dens-temp-nc}
\end{figure*}

Figure~\ref{fig:im-dens-temp-nc} shows the equivalent information as Figure~\ref{fig:im-dens-temp}, but now for the non-colliding simulation. These clouds generally have lower mass surface densities, mass-weighted densities and mass-weighted temperatures than are seen in the colliding case. Again, application of a density threshold is important to isolate the clouds from the ambient medium.

\begin{figure*}
    \centering
    \includegraphics[width=\linewidth]{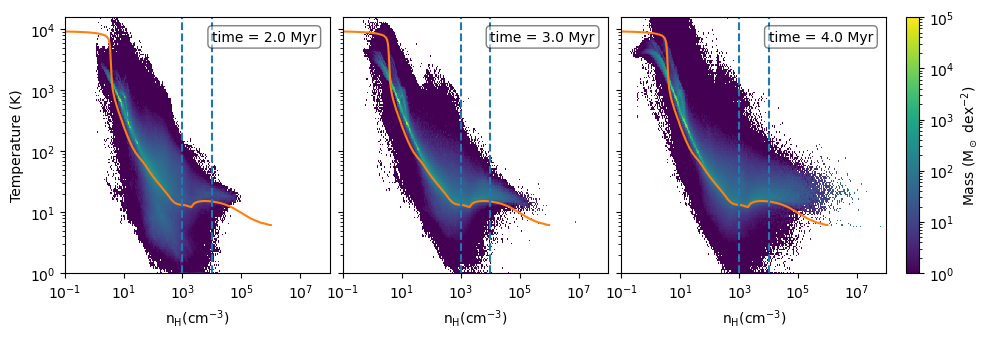}
    \includegraphics[width=\linewidth]{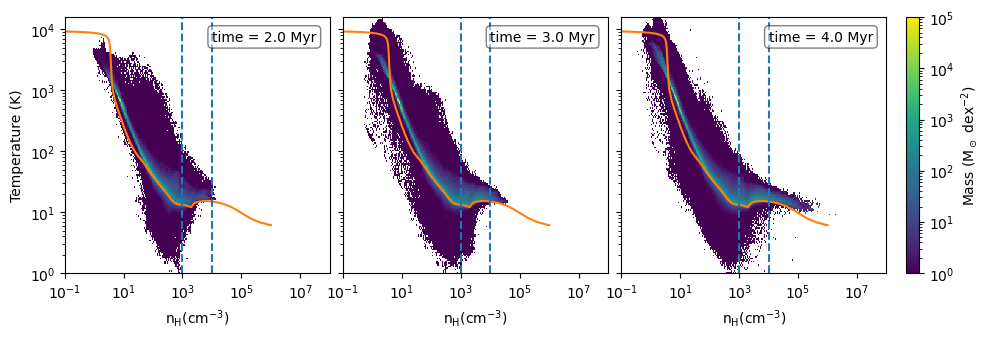}
    \caption{Two-dimensional mass distributions of density and temperature. The first row shows the results from the colliding case and the second shows the results from the non-colliding case. The color indicates the mass per unit area in the phase space diagram of ${\rm log}\:n_{\rm H}$ and temperature. The orange curve shows the equilibrium temperature from the heating and cooling functions. The blue dashed lines indicate the densities of $n_{\rm H} = 10^3$ and $10^4\:{\rm cm}^{-3}$.}
    \label{fig:hist2d-dens-temp}
\end{figure*}

We also examine the relation between density and temperature in the cells. Figure~\ref{fig:hist2d-dens-temp} shows the density and temperature phase space distribution of the colliding case in the $n_{\rm H} - {T}$ plane. The equilibrium temperature from the heating and cooling functions is overplotted with a solid line. We notice that the densities are usually lower than $10^5\:{\rm cm}^{-3}$ at 2~Myr. Even at 3~Myr, there is relatively little material at densities higher than $10^5\:{\rm cm}^{-3}$. Thus, we see that the short lifetime of the high density regions (up to 4~Myr) justifies the usage of the reduced model to only have a good agreement with the `Grid 2' model of \citet{Entekhabi2022} at high densities for the limited period up to 1~Myr (see \S\ref{sec:method}). We also notice that at a given density, the temperature shows a distribution approximately centered about the equilibrium temperature for densities up to about $n_{\rm H} = 10^4\:{\rm cm}^{-3}$. However, at higher densities the temperature appears relatively elevated, i.e., at about 20~K, compared to the equilibrium value, which is near 10~K. \citetalias{Hsu2023} concluded this elevated temperature is likely due to compressional heating, also noting that the dense gas in the non-colliding clouds is cooler and closer to the values expected from simple thermal equilibrium.

\subsection{C and O reservoirs\label{sec:coreservoir}}

\begin{figure*}
    \centering
    \includegraphics[width=\linewidth]{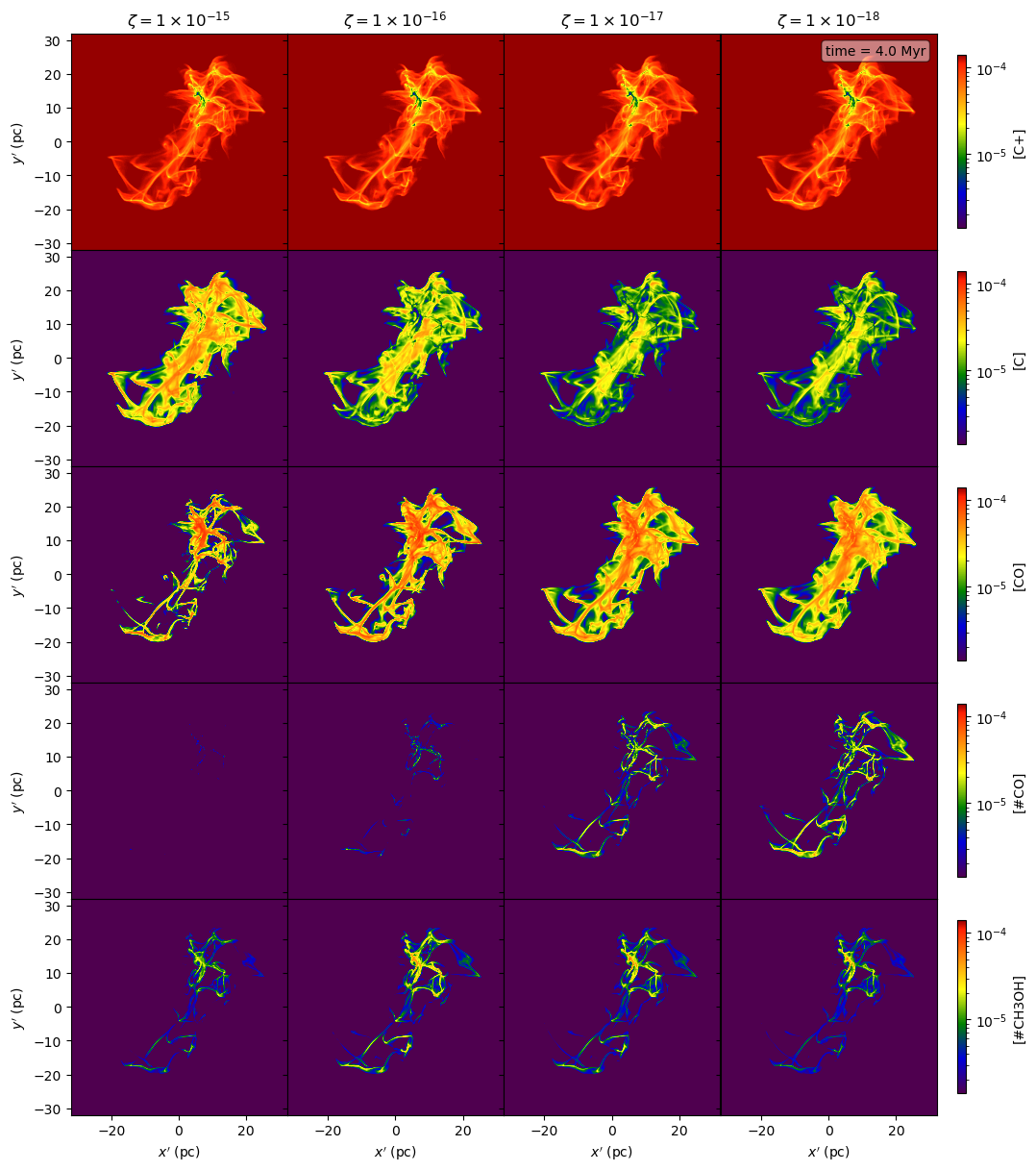}
    \caption{The rows from top to bottom show maps of $\rm C^+$, C, CO, ice-phase CO and ice-phase $\rm CH_3OH$ of the colliding case at 4~Myr. The columns from left to right show the results under different CRIRs ($\zeta = 10^{-15}$, $10^{-16}$, $10^{-17}$ and $10^{-18}\:{\rm s^{-1}}$). }
    \label{fig:cc-im-cp-c-co-gch3oh}
\end{figure*}

\begin{figure*}
    \centering
    \includegraphics[width=\linewidth]{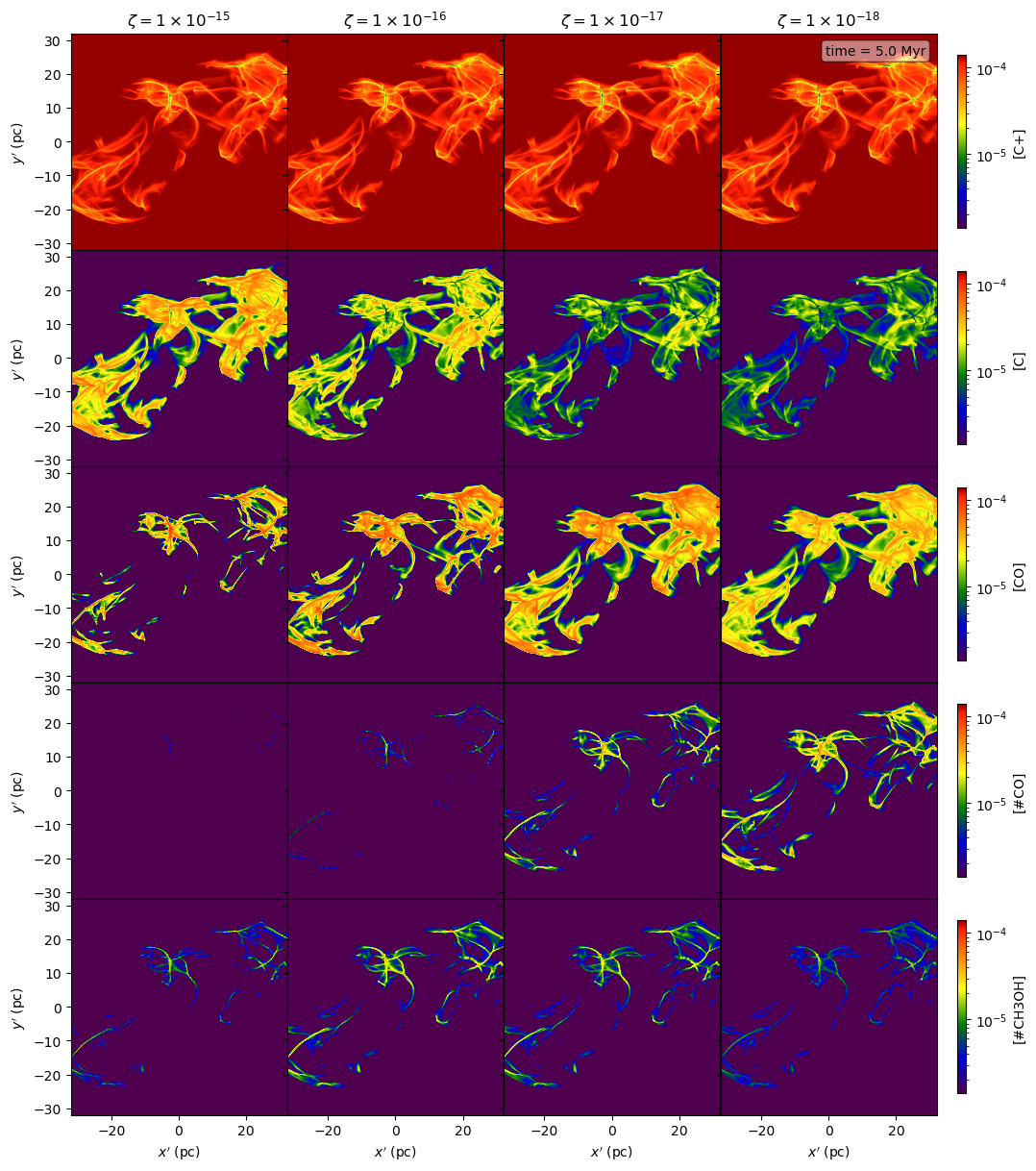}
    \caption{As Figure~\ref{fig:cc-im-cp-c-co-gch3oh}, but for the non-colliding case at 5~Myr.}
    \label{fig:nc-im-cp-c-co-gch3oh}
\end{figure*}

Figures~\ref{fig:cc-im-cp-c-co-gch3oh} and \ref{fig:nc-im-cp-c-co-gch3oh} show maps of abundances of $\rm C^+$, C, CO, \#CO and \#$\rm CH_3OH$ (i.e., including the ice-phases of CO and $\rm CH_3OH$, following the notation used in UCLCHEM) in the colliding and non-colliding cases, respectively. In both we see that the dominant C reservoir is $\rm C^+$ in the ambient gas, with the carbon kept ionized by the uniform radiation field $F_{\rm UV} = 4G_0$. However, as density (and thus $A_V$) increases, atomic C becomes more dominant. As density increases further, C is converted into CO. The thickness and importance of the C layer increase with CRIR, as also found by \citet{2021MNRAS.502.2701B}. We also note that the morphology of the CO map has a similar boundary to the mass surface density map identified with a density threshold of $n_{\rm H} = 10^3\:{\rm cm^{-3}}$. This indicates the ability of CO to be a tracer of the molecular cloud: in the lower density envelopes of the GMC, CO is dissociated, i.e., these are regions of ``CO-dark'' molecular gas. In the densest, coldest regions, CO begins to freeze-out onto dust grains, i.e., we see a rise in the abundance of CO ice. Note that our adopted chemical assumes that 90\% of CO forms \#CO and 10\% forms \#$\rm CH_3OH$ when CO sticks on dust grains. As a result of this assumed branching ratio and the cycling of CO back to the gas phase by thermal and cosmic ray induced desorption processes, the global amount of C in $\rm CH_3OH$ ice increases over time to relatively high values.




\begin{figure*}
    \centering
    \includegraphics[width=\linewidth]{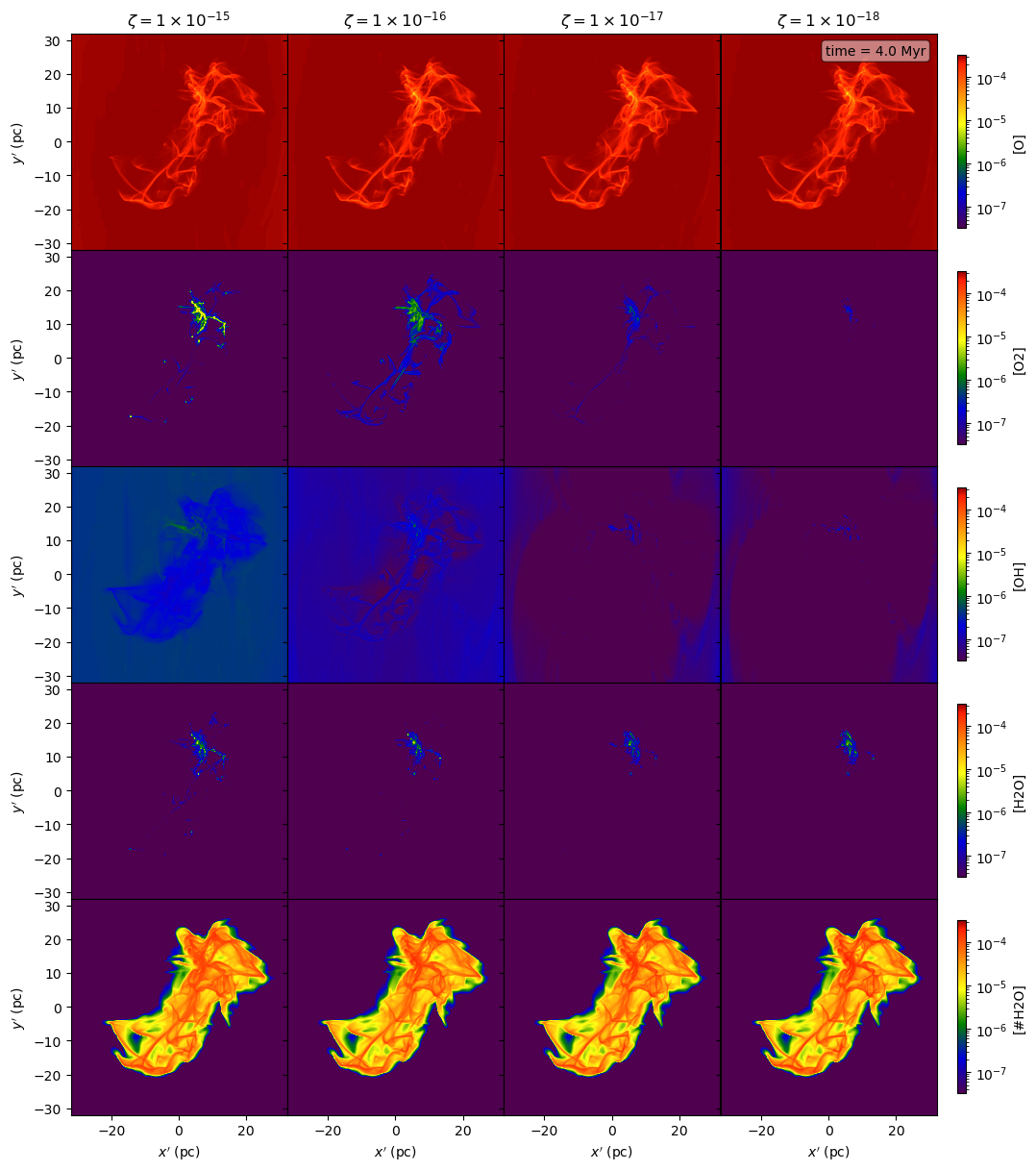}
    \caption{The rows from top to bottom show the maps of O, $\rm O_2$, OH, water and water ice of the colliding case at 4~Myr. The columns from left to right show the results under different CRIRs ($\zeta = 10^{-15}$, $10^{-16}$, $10^{-17}$ and $10^{-18}\:{\rm s^{-1}}$). }
    \label{fig:cc-im-o-o2-gh2o}
\end{figure*}

\begin{figure*}
    \centering
    \includegraphics[width=\linewidth]{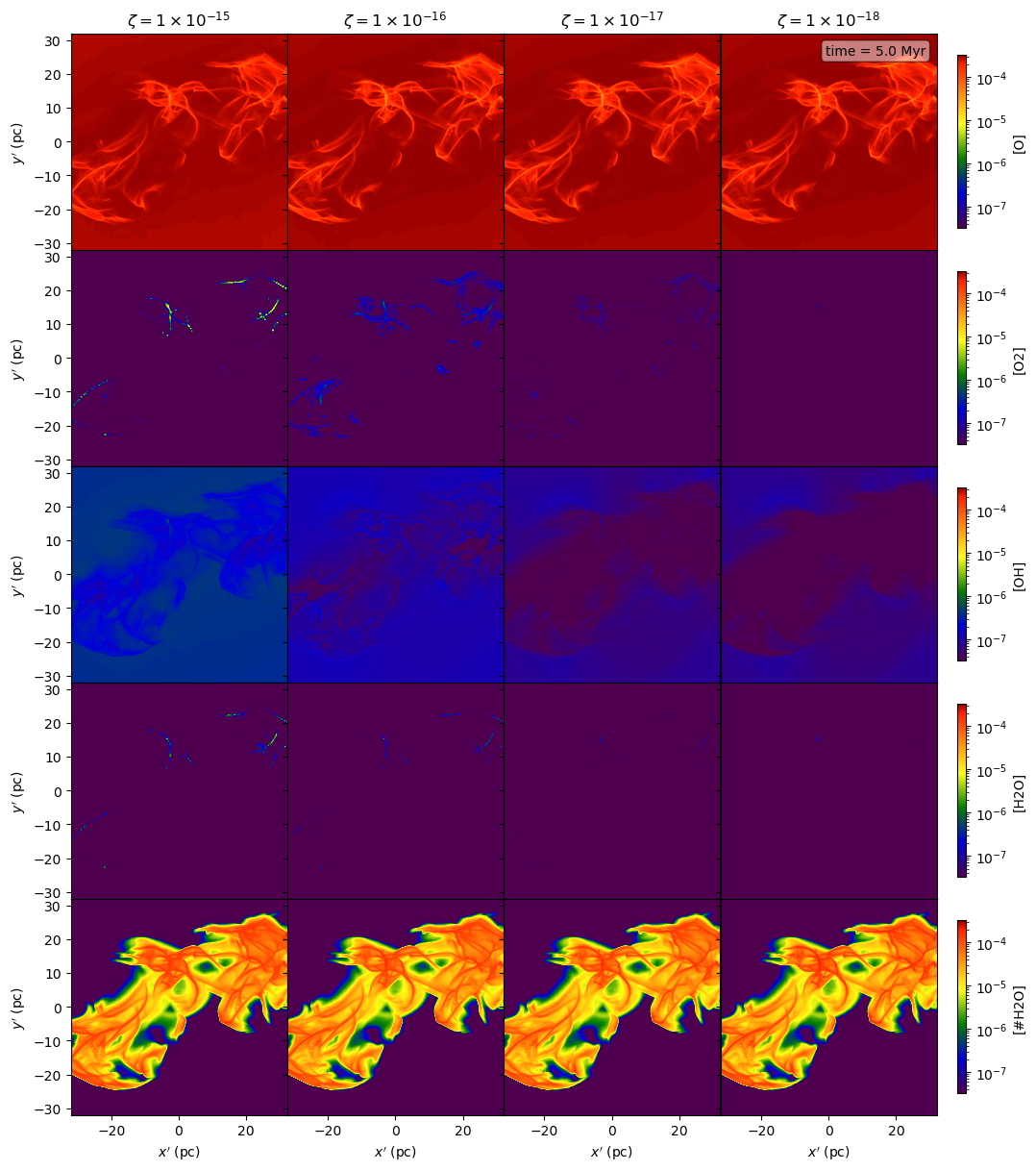}
    \caption{As Figure~\ref{fig:cc-im-o-o2-gh2o}, but for the non-colliding case at 5~Myr.}
    \label{fig:nc-im-o-o2-gh2o}
\end{figure*}

We also examine the O reservoir in Figures~\ref{fig:cc-im-o-o2-gh2o} and \ref{fig:nc-im-o-o2-gh2o}. At low densities, most O stays in atomic form. In denser regions, the remaining O that is not in CO is mostly incorporated into water ice on grain surfaces. Modest amounts of $\rm O_2$ are seen to form in some of the densest regions, especially when they are subject to higher CRIRs. Gas phase water is also present in localized dense regions with abundances of $\sim 10^{-7}$ to $\sim 10^{-5}$. Its morphology remains relatively constant for CRIRs $\lesssim 10^{-16}\:{\rm s}^{-1}$, so we attribute its presence to the effects of shock heating of dense gas and subsequent thermal desorption from grain ice mantles.


\begin{figure*}
    \centering
    \includegraphics[width=\linewidth]{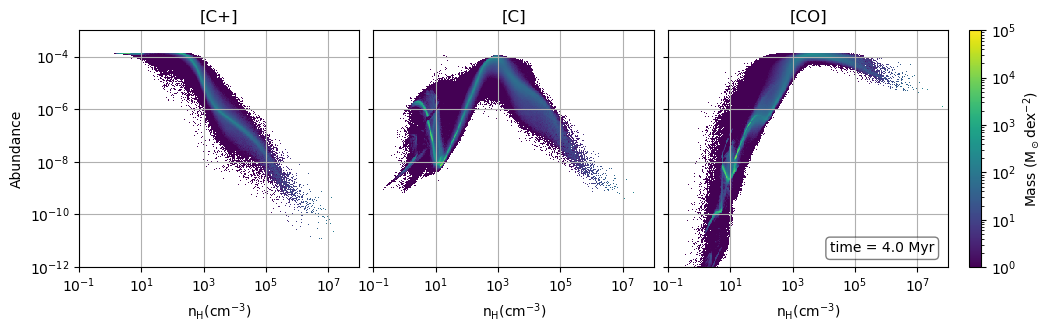}
    \includegraphics[width=\linewidth]{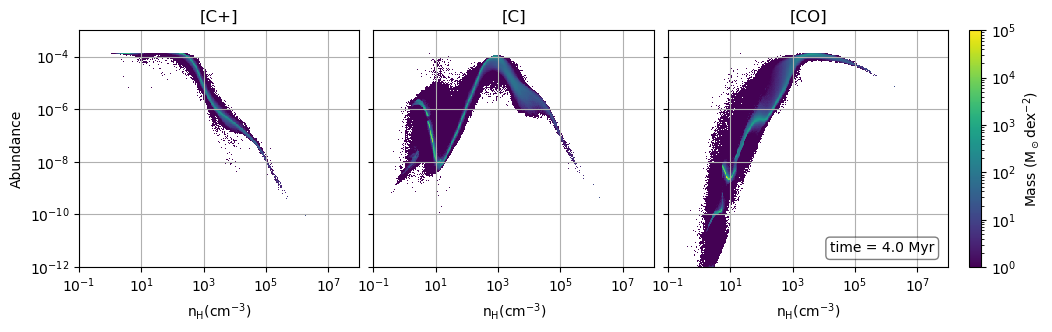}
    \caption{The density-abundance phase plots of $\rm C^+$, C and CO of the fiducial, i.e., $\zeta=10^{-16}\:{\rm s}^{-1}$, colliding (top) and non-colliding cases (bottom) at 4~Myr. }
    \label{fig:hist2d-cp-c-co}
\end{figure*}

Figure~\ref{fig:hist2d-cp-c-co} shows abundance versus density phase plots for the main C gas phase species of the fiducial case ($\zeta = 10^{-16}\:{\rm s^{-1}}$). This reveals more details about the transitions from $\rm C^+$ to C to CO. We see that $\rm C^+$ is the dominant C reservoir if the density is lower than $n_{\rm H} \sim 300\:{\rm cm^{-3}}$. Its abundance then starts to decrease at higher densities, balanced by an increase in atomic C. Atomic C is dominant over a relatively narrow range of densities near $n_{\rm H} \sim 1,000\:{\rm cm^{-3}}$. At higher densities, CO takes over as being the main reservoir of C. At densities approaching $n_{\rm H} \sim 10^5\:{\rm cm^{-3}}$, we see a modest decrease in gas phase CO abundance due to formation of CO ice and $\rm CH_3OH$ ice on dust grains. In the next sub-section we examine this depletion of CO into ice species in more detail by focusing on localized high density regions.

\subsection{High Density Regions\label{sec:dense-gas}}

\begin{figure*}
    \centering
    \includegraphics[width=\linewidth]{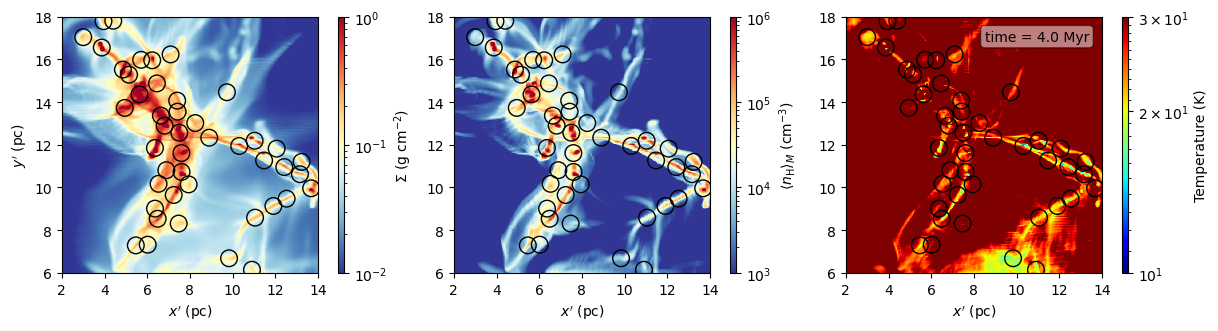}
    \includegraphics[width=\linewidth]{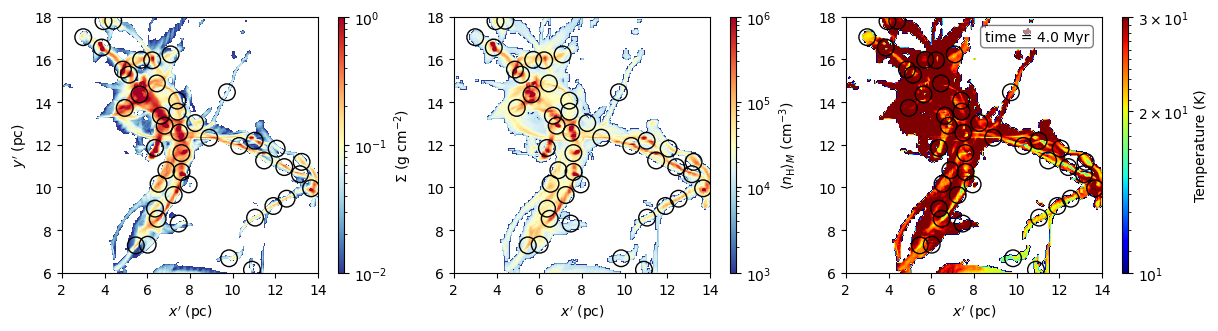}
    \caption{
    High density region ($12\times12\times12\:$pc) formed from a GMC-GMC collision. Maps of mass surface density (left column), mass-weighted number density (middle column) and mass-weighted temperature (right column) are shown. The top row shows the results of the fiducial case ($\zeta = 10^{-16}\rm s^{-1}$), while the second row shows the same region with a density threshold of $n_{\rm H}=10^4\:{\rm cm}^{-3}$ used to select cells. } 
    \label{fig:cc-im-zoom-dens-temp}
\end{figure*}

Here, we focus on the chemical evolution in some example relatively high density regions. We will see that these regions are most comparable to conditions that are found in IRDCs, i.e., the early stages of star-forming clumps that will eventually become star clusters. From the colliding case, we select a $12\times12\times12\:$pc box that contains some of the densest structures formed at a time of 4~Myr. Figure~\ref{fig:cc-im-zoom-dens-temp} shows the projected mass surface density, mass-weighted number density and mass-weighted temperature of this high density region. We show a case where no density threshold is applied (top row) and a case where only cells above a density threshold of $n_{\rm H}=10^4\:{\rm cm}^{-3}$ contribute to the map.
Note, that for the high density region the projection is only done through a depth of 12~pc, so a large part of the ambient gas from outside the initial GMCs is automatically excluded. This reduces the importance of a density threshold on the mass-weighted quantities, especially for the densest clumps.

From Figure~\ref{fig:cc-im-zoom-dens-temp}, we see that there is a main high mass surface density structure that is approximately filamentary and that extends in a curved arc from north to south over a length of about 10~pc. It is joined by a thinner filamentary spur running east-west. The peak values of $\Sigma$ reach to $\sim 1\:{\rm g\:cm}^{-2}$. The mass-weighted number density map reveals most clearly that the main filament has fragmented into about 10 major clumps. The maximum number densities achieved are $n_{\rm H}\sim10^6\:{\rm cm}^{-3}$. The mass-weighted temperatures reach minimum values of around 20~K, but we see there are global temperature gradients in the dense clumps, i.e., being relatively colder in the southern region of the main filament. 



To further compare with observational data, we select some sample clumps from the mass surface density map (without density threshold) using the dendrogram algorithm. With the {\tt Astrodendro} \citep{Rosolowsky2008} python package, we first find the local density peaks on the mass surface density map. Then, at the center of each density peak, we select a circle with a radius of 0.39~pc. This radius is the same as that of the half power beam width (HPBW) of IRAM (Instituto de Radioastromía Milimétrica) 30m telescope for a source at a 5~kpc distance, which is the configuration used for the observational data in \citet{Entekhabi2022}. The parameters used to find the density peaks are similar to \citetalias{Hsu2023}, which used minimum mass surface density $\Sigma_{\rm min} = 0.1\,{\rm g\, cm^{-2}}$, minimum increment of mass surface density $\delta_{\rm min} = 0.025\,{\rm g\, cm^{-2}}$. However, we use a slightly larger value of the minimum area $A_{\rm min} = 16$ pixels in this study. This larger size of the clumps reduces the overlap among cores when we then select regions with circular apertures of 0.39~pc radius. With this approach, we identify about 25 clumps in the colliding case at 2~Myr and the number increases to about 40 at 4~Myr. We note that the positions of the clumps are fixed once we find them on the mass surface density map. Then subsequent analysis that applies different density thresholds is done for the same defined clump regions.

\begin{figure*}
    \centering
    \includegraphics[width=\linewidth]{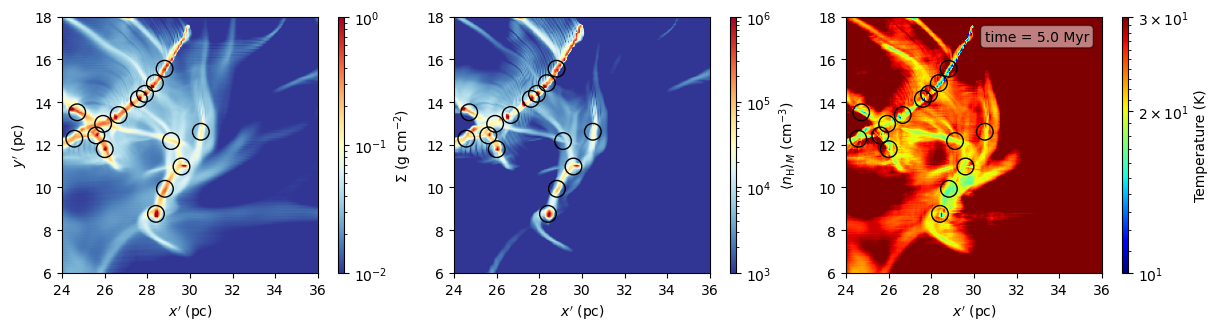}
    \includegraphics[width=\linewidth]{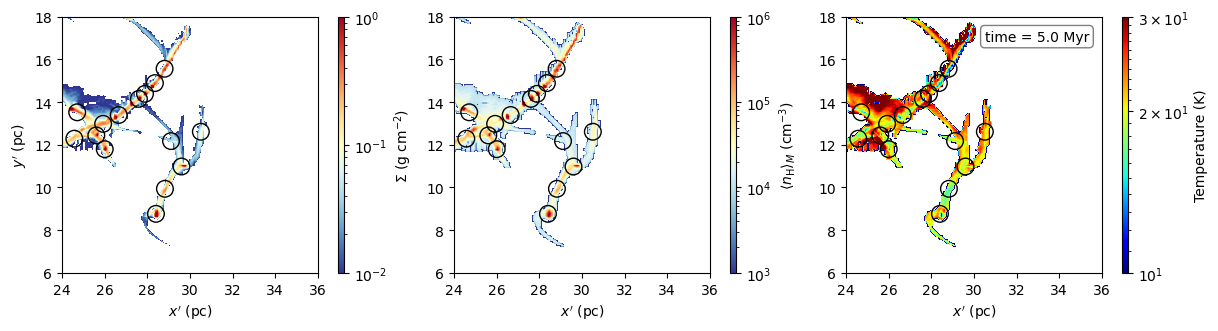}
    \caption{Maps of mass surface density (left column), mass-weighted density (middle column) and mass-weighted temperature (right column), with black circles indicating the selected clumps, for the non-colliding case at 5~Myr.}
    \label{fig:nc-im-zoom-dens-temp}
\end{figure*}

For the non-colliding case, which takes longer to develop dense structures, we select the dense region at 5~Myr. Figure~\ref{fig:nc-im-zoom-dens-temp} shows this dense region, i.e., it is taken from near the center of one of the GMCs. As has been noted in previous papers in this series, overall compared to the colliding case, the non-colliding simulation produces smaller quantities of dense gas during the evolution of the clouds, even if this is extended to 5~Myr. The region contains one main filament, 8~pc in length, that is relatively thin and straight compared to the main filament in the colliding case. A secondary, smaller filament is also present. The regions of high mass surface density are smaller and peak at more modest values compared to the colliding case. Similarly, peak number densities are lower, $\sim {\rm few}\times 10^5\:{\rm cm}^{-3}$. We also notice that temperatures in the dense gas are significantly cooler than in the colliding case, with some regions approaching temperatures of $\sim 15\:$K. Applying the same dendrogram algorithm to find dense clumps, we obtain a sample of 7 objects at 3~Myr and about 15 at 5~Myr, i.e., about one third of the number found in the colliding case at 2~Myr and 4~Myr.

\subsubsection{$CO$ Depletion Factor}

Here we examine the freeze-out, i.e., ``depletion'', of CO from the gas phase in the high density regions. This depletion to form CO ice is generally expected to become efficient in dense gas that cools down to temperatures of $\lesssim20\:$K \citep{Caselli1999a}. Observationally, CO depletion is typically measured via the CO depletion factor, i.e., $f_D$(CO) \citep[e.g.,][]{2011ApJ...738...11H}, which is defined as:
\begin{equation}
    f_{D} = \frac{[{\rm CO}]_{\rm ideal}}{[{\rm CO}]_{\rm gas}} = \frac{1.4 \times 10^{-4}}{[{\rm CO}]_{\rm gas}},
\end{equation}
where $[{\rm CO}]_{\rm ideal}$ is an idealised CO abundance assuming all C is in the form of gas-phase CO and $[{\rm CO}]_{\rm gas}$ is the observed gas-phase abundance of CO. Note, observational studies of $f_D$(CO) are based on: (1) projected maps of total mass surface density, which set the idealised column density of CO; (2) integrated intensity (moment zero) maps over a defined velocity range of CO line emission, which yield a measurement of the observed gas-phase abundance. Of course, within our simulation domain, we also have access to the 3D structure of $f_D$(CO) that can be measured in each cell and can also make direct measures of the column density of CO from a defined region.

However, because of the possibility that some C is in the form of atomic C or C$^+$ and that some ice-phase C is in the form of species other than CO, especially, in our model, ice-phase $\rm CH_3OH$, we also consider two other ways to measure the fraction of CO that is depleted onto dust grains. One way is to measure the ratio of total CO (gas-phase and ice-phase) to gas-phase CO, i.e.:
\begin{equation}
    f_{\rm D,1} = \frac{[{\rm CO}]_{\rm gas}+[{\rm CO}]_{\rm ice}}{[{\rm CO}]_{\rm gas}}
\end{equation}
Another way is similar, but replaces the ice-phase CO to the total ice-phase C, i.e.:
\begin{equation}
    f_{\rm D,2} = \frac{[{\rm CO}]_{\rm gas}+[{\rm C-species}]_{\rm ice}}{[{\rm CO}]_{\rm gas}}.
\end{equation}



\begin{figure*}
    \centering
    \includegraphics[width=0.9\linewidth]{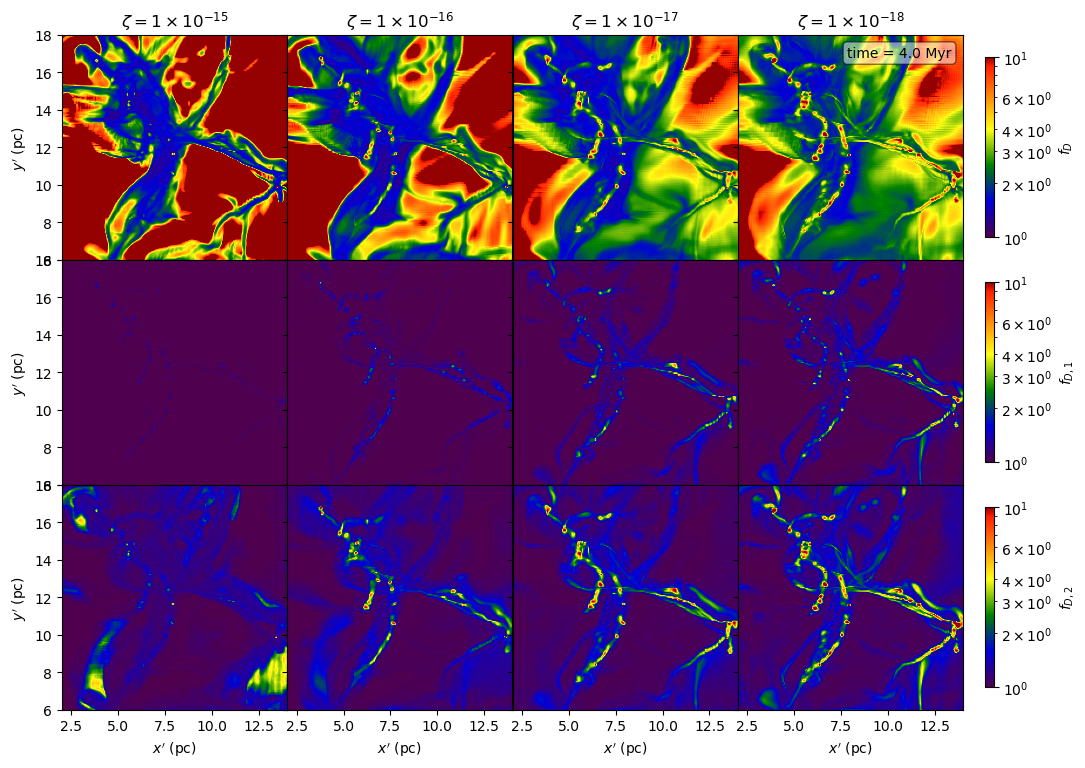}
    \includegraphics[width=0.9\linewidth]{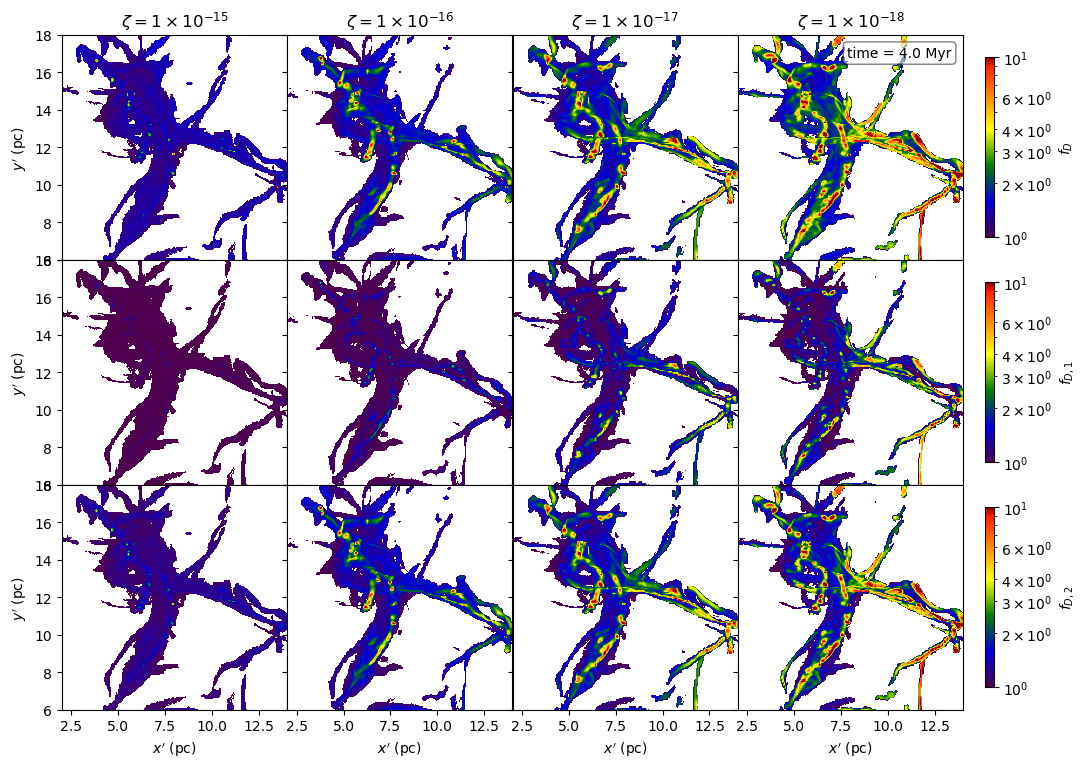}
    \caption{Maps of different CO depletion factors of the colliding case at 4~Myr. The top panels show the CO depletion factors from the whole 12~pc$^3$ box, while the bottom panels include a density threshold of $10^4\:{\rm cm}^{-3}$. The rows in each panel show $f_{D}, f_{D,1}, f_{D,2}$ from top to bottom (see text). The columns from left to right show the results under different CRIRs ($\zeta = 10^{-15}$, $10^{-16}$, $10^{-17}$ and $10^{-18}\:{\rm s^{-1}}$).}
    \label{fig:cc-im-codep}
\end{figure*}

Figure~\ref{fig:cc-im-codep} shows the maps of the three CO depletion factors for the high density region of the colliding case and the impact of difference CRIRs. Similar to Figure~\ref{fig:cc-im-zoom-dens-temp}, we also consider maps without and with a density threshold (of $n_{\rm H} = 10^4\:{\rm cm}^{-3}$). From the first group of panels without any density threshold, we see that the CO depletion factor is very high outside the dense gas due to the dominance of $\rm C^+$. With weaker CRIR, some spots of high deplection factor appear along with the arc shown in Figure~\ref{fig:cc-im-zoom-dens-temp}. The comparison with the $f_{D, 1}$ and $f_{D, 2}$ shows that those local peaks are contributed more by other ice-phase C-bearing species than by CO. From the maps from the high density region considering density threshold, we can further confirm that the standard CO depletion factor maps are significantly different from the $f_{D,1}$ maps, but quite similar to the $f_{D,2}$ maps. The comparison indicates that those ice-phase C-bearing species reside inside the arc and they are the main contributor to the CO depletion factors, i.e., especially methanol given our UCLCHEM model assumptions. 
We see in the colliding case, e.g., when a density threshold has been applied, that when the CRIR is $10^{-17}\:{\rm s}^{-1}$ or lower, the CO depletion factors $f_D$ or $f_{D,2}$ reach peak values $\sim 10$ or greater in some of the dense clumps, especially in the colder southern clumps in the main filament and those in the western spur. For CRIRs $\sim 10^{-16}\:{\rm s}^{-1}$ the peak CO depletion factors are $\sim 5$. There is very little CO depletion at the higher CRIR considered of $10^{-15}\:{\rm s}^{-1}$.

\begin{figure*}
    \centering
    \includegraphics[width=0.9\linewidth]{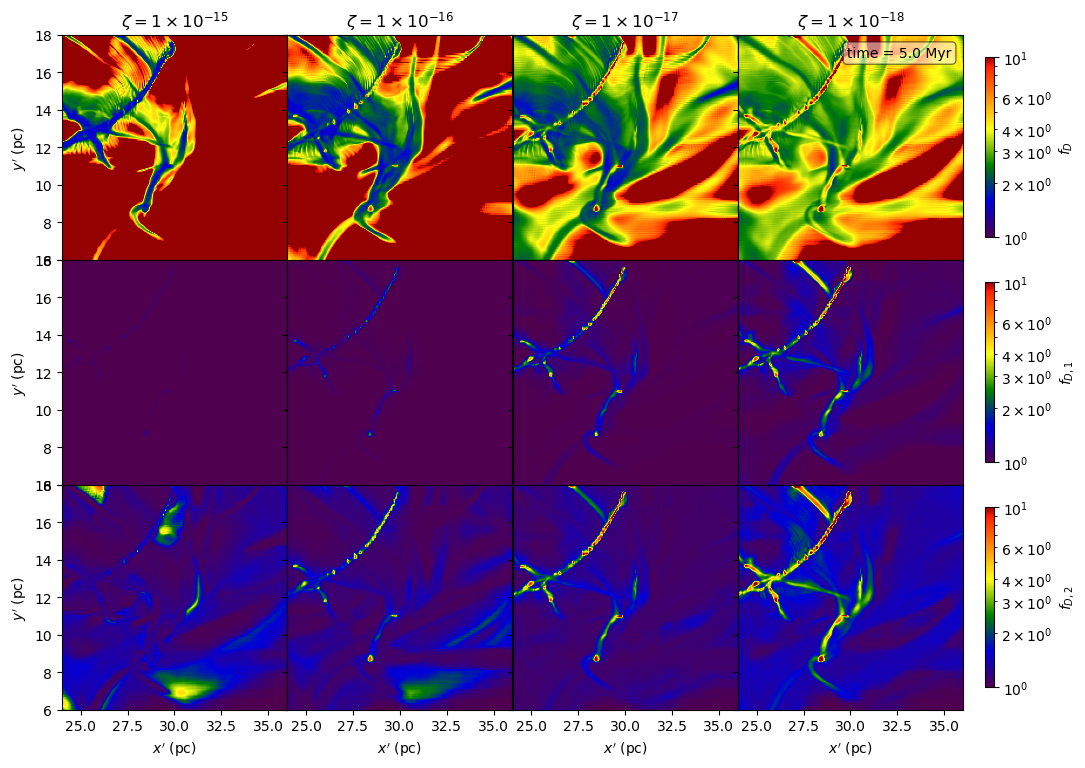}
    \includegraphics[width=0.9\linewidth]{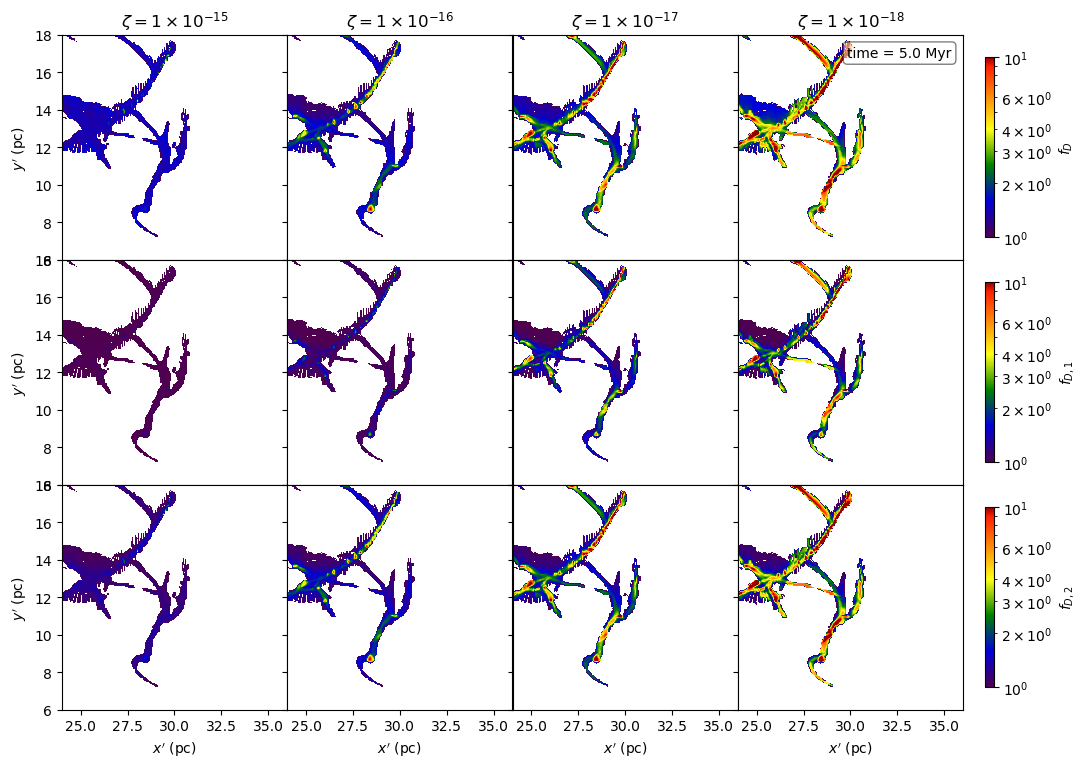}
    \caption{As Figure~\ref{fig:cc-im-codep}, but for the non-colliding case at 5~Myr.}
    \label{fig:nc-im-codep}
\end{figure*}

Figure~\ref{fig:nc-im-codep} shows the equivalent CO depletion factor maps for the non-colliding case. We find similar behaviour as in the colliding case. The main filament is relatively hard to distinguish in the $f_D$ map, unless a density threshold is applied. However, we also see relatively high CO depletion factors in this filament, which has a relatively cool temperature near 15~K, when the CRIR is $10^{-17}\:{\rm s}^{-1}$ or lower. We note that the colliding case has more regions of higher density, which promotes CO depletion, but has moderately warmer temperatures, which tends to reduce the efficiency of freeze-out.

\begin{figure*}
    \centering
    \includegraphics[width=\linewidth]{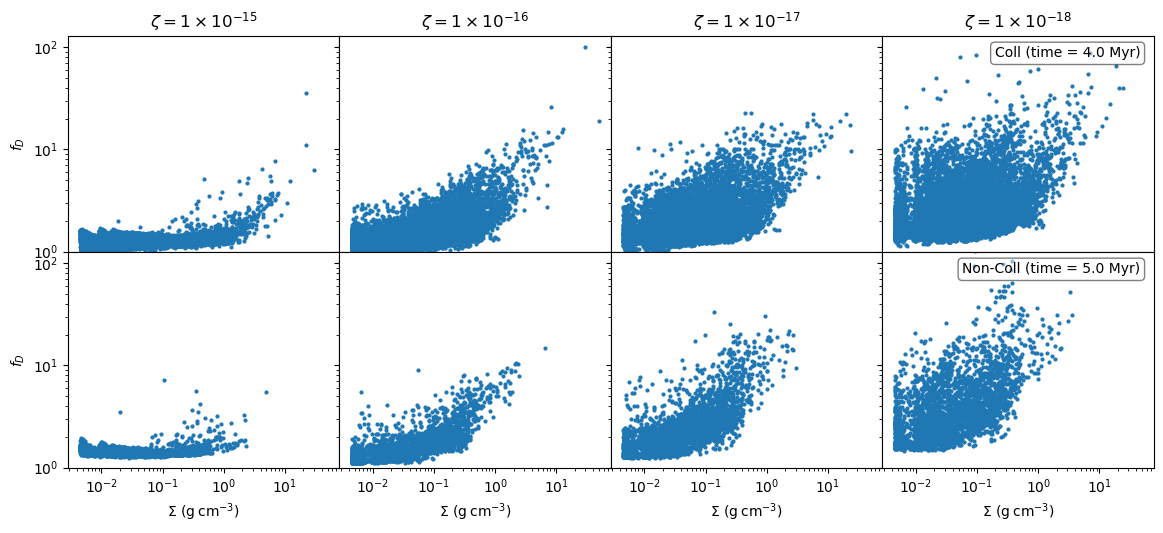}
    \includegraphics[width=\linewidth]{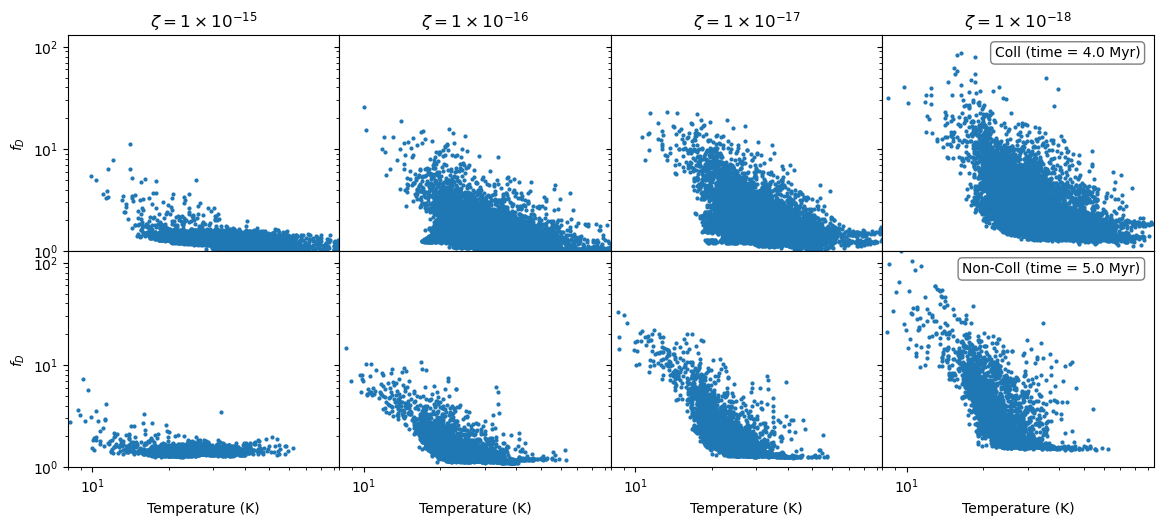}
    \caption{{\it (a) Top panel:} Scatter plots of CO depletion factor ($f_D$(CO)) versus mass surface density derived from the depletion factor maps (including density threshold of $n_{\rm H} = 10^4\:{\rm cm}^{-3}$; see Figures~\ref{fig:cc-im-codep} and \ref{fig:nc-im-codep}) in the colliding case at 4~Myr (upper row) and non-colliding case at 5~Myr (lower row), for various CRIRs (columns, as labelled).
    {\it (b) Bottom panels:} As (a), but now showing CO depletion factor versus mass-weighted temperature.
    }
    \label{fig:cc-nc-sc-codep}
\end{figure*}

By considering the pixels of the CO depletion factor maps shown in Figures~\ref{fig:cc-im-codep} and \ref{fig:nc-im-codep} with a density threshold of $n_{\rm H} = 10^4\:{\rm cm}^{-3}$, we show scatter plots illustrating the dependence of $f_D$(CO) on mass surface density and temperature for the colliding and non-colliding simulations and for our explored range of CRIRs in Figure~\ref{fig:cc-nc-sc-codep}. Here we see the suppression of CO depletion at high CRIRs. These properties can be observationally measured in molecular clouds, especially IRDCs \citep[e.g.,][]{2011ApJ...738...11H, Entekhabi2022}, so there is the potential to constrain the CRIR and/or the efficiency of cosmic ray induced desorption processes via the comparison with such data.

\subsubsection{$CO$, $HCO^+$, \& $N_2H^+$}
\label{sec:co-hco-n2h}

Here we examine the abundances of three simple molecules, CO, $\rm HCO^+$ and $\rm N_2H^+$, which are typical dense gas tracers. These were also the main species used in the study of \citet{Entekhabi2022} to constrain the chemical age and environmental conditions in 10 IRDC clumps. The abundances are obtained by the ratio of species column densities and the total column density of H nuclei, i.e., $N_{\rm H}$.

\begin{figure*}
    \centering
    \includegraphics[width=0.9\linewidth]{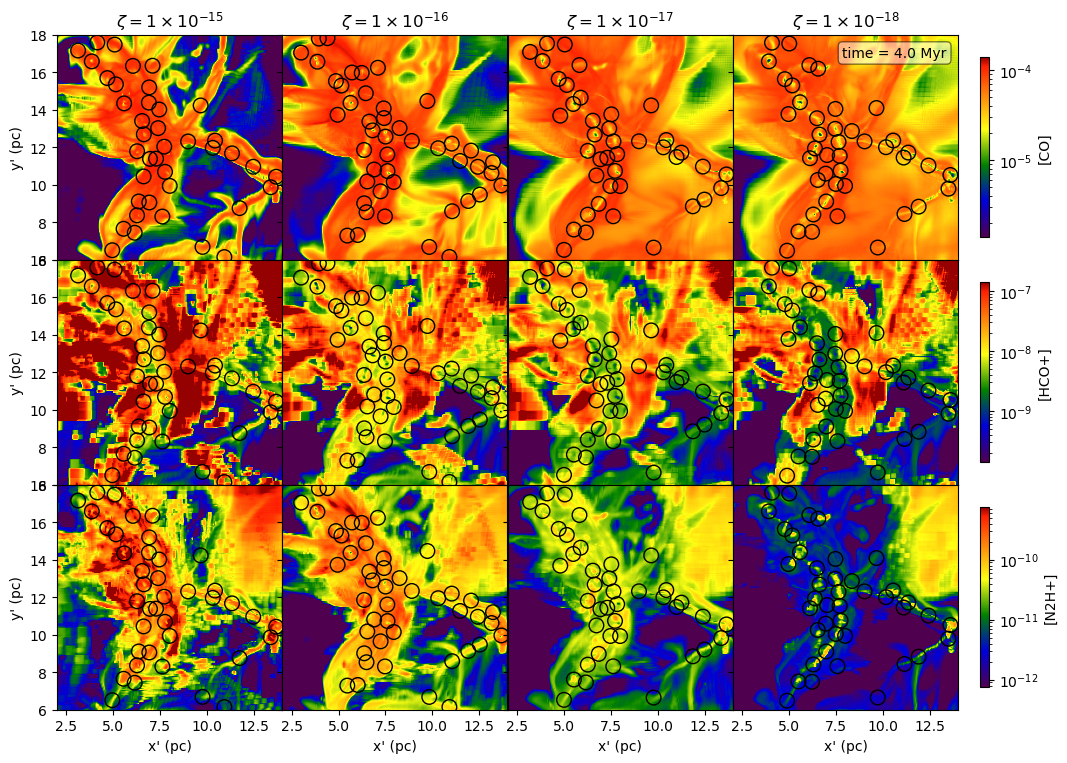}
    \includegraphics[width=0.9\linewidth]{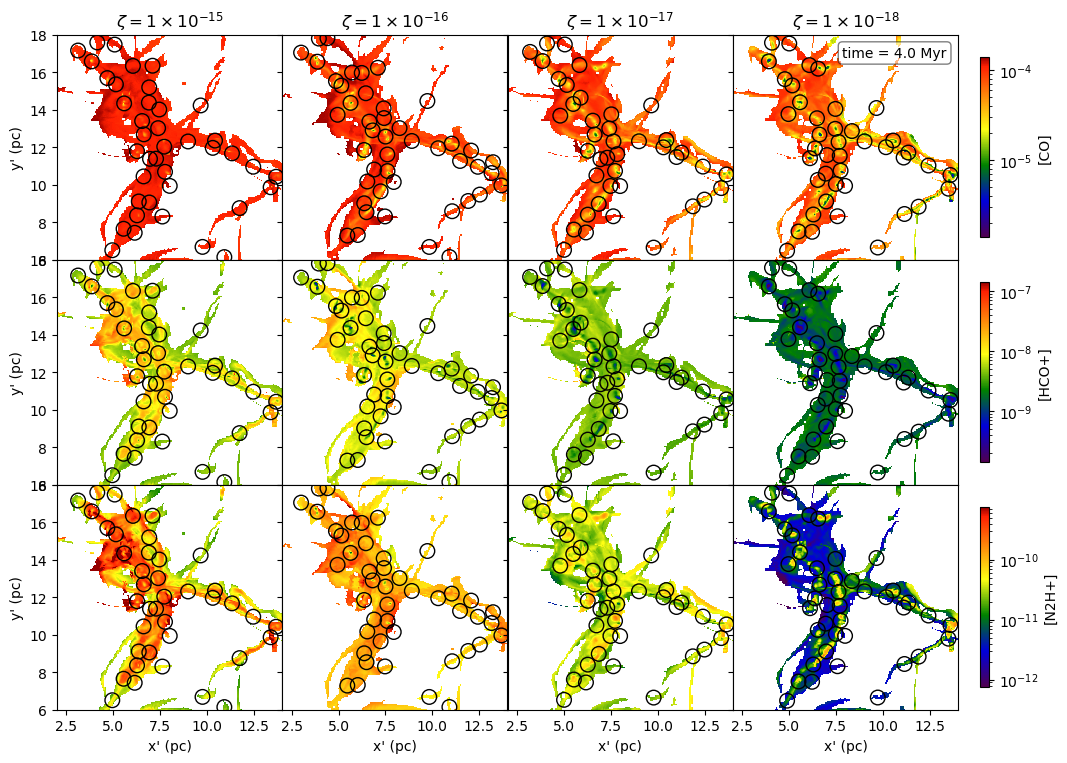}
    \caption{Top panel: maps of CO (first row), ${\rm HCO^+}$ (second row) and ${\rm N_2H^+}$ (bottom row) for the colliding case at 4~Myr. The columns from left to right indicate results for different CRIRs: ($\zeta = 10^{-15}$, $10^{-16}$, $10^{-17}$ and $10^{-18}\:{\rm s^{-1}}$).
    Bottom panel: Same as top panel, but now with a density threshold of $n_{\rm H}=10^4\:{\rm cm}^{-3}$ for inclusion in the map.
    }
    \label{fig:cc-im-co-hcop-n2hp}
\end{figure*}

\begin{figure*}
    \centering
    \includegraphics[width=0.9\linewidth]{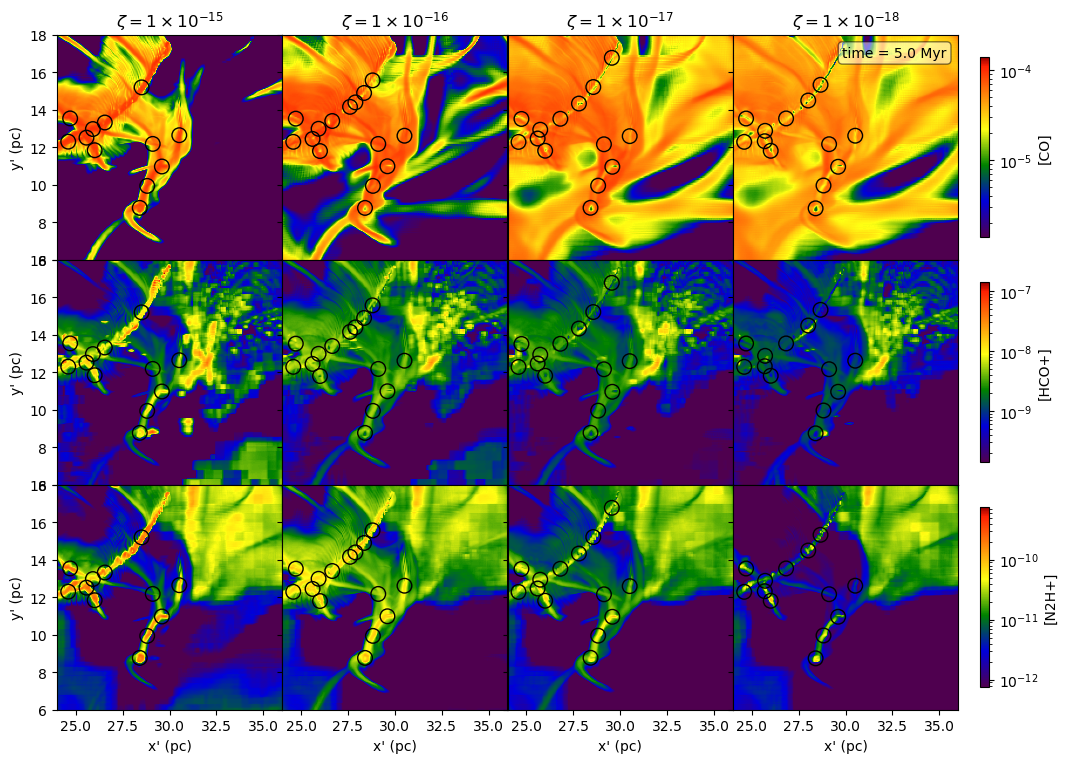}
    \includegraphics[width=0.9\linewidth]{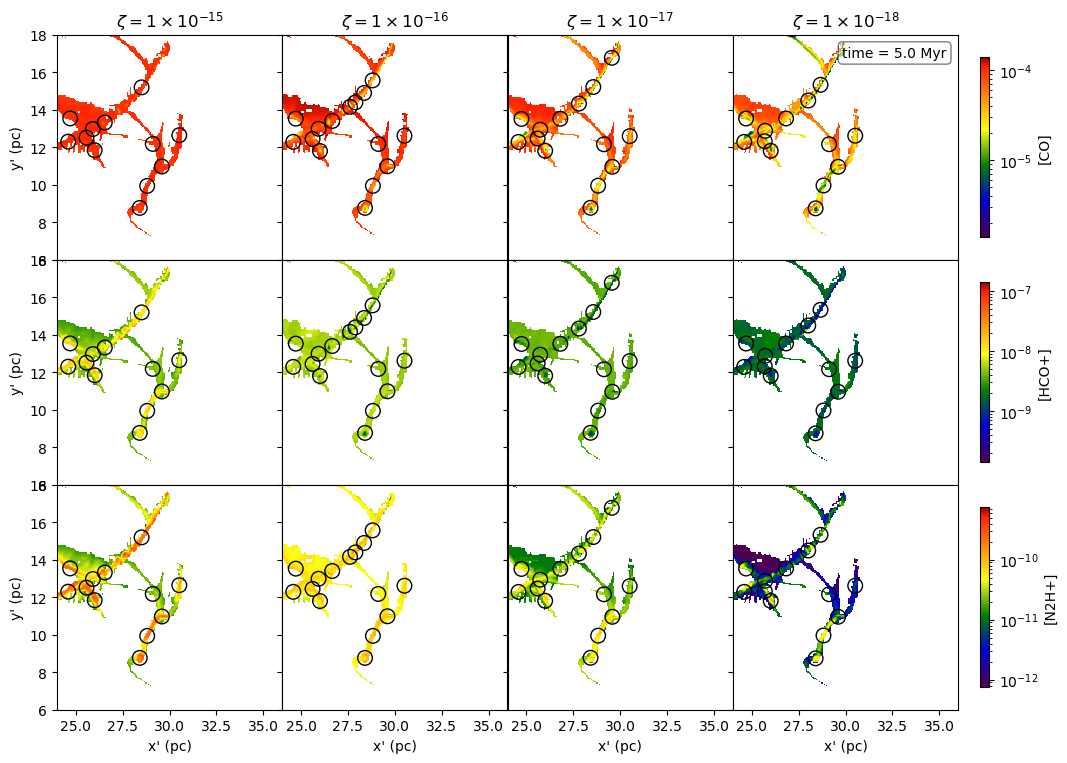}
    \caption{As Figure~\ref{fig:cc-im-co-hcop-n2hp}, but now for non-colliding case at 5~Myr.}
    \label{fig:nc-im-co-hcop-n2hp}
\end{figure*}

Figure~\ref{fig:cc-im-co-hcop-n2hp} and \ref{fig:nc-im-co-hcop-n2hp} show the maps of CO, $\rm HCO^+$, and $\rm N_2H^+$ abundances from the same high density regions of the colliding case and the non-colliding case, respectively. Again, we show both the influence of density threshold and CRIR. In Section~\ref{sec:coreservoir}, the large-scale maps have shown that higher CRIRs suppress the region of CO formation. In the high density region maps without density threshold, we see again that the CO is confined inside the arc when $\zeta = 10^{-15}\:{\rm s}^{-1}$ and the effect of CO depletion is hard to be seen. Instead, in the weakest CRIR, we see a widespread CO abundance and CO depletion along the filamentary structures, as we have found in Figure~\ref{fig:cc-im-codep}.

In the maps of $\rm HCO^+$, we first see that higher CRIRs induce globally higher abundances of $\rm HCO^+$. In addition, we find that $\rm HCO^+$ is more abundant outside the densest regions. This is partly explained by the freeze out of CO in dense regions, but also because of the dependence on recombination reactions that tend to reduce ionization levels in denser regions. We also note that observationally $\rm HCO^+$ has high critical densities for its rotational transition lines, e.g., $\sim 1.80 \times 10^5\:\rm cm^{-3}$ for $\rm HCO^+$(1-0) at 15~K according to the LAMDA database. Therefore, the lower density regions may be less efficient at producing strong emission. To examine this, we present synthetic line emission maps in \S\ref{sec:obs}.

The maps of $\rm N_2H^+$ also show an increasing abundance as CRIR is raised. We note that $\rm N_2H^+$(1-0) rotational transition also has similar high critical densities, i.e., $\sim 1.58 \times 10^5\:\rm cm^{-3}$ at 15~K, as $\rm HCO^+$. A difference between the maps of $\rm HCO^+$ and $\rm N_2H^+$ is that $\rm N_2H^+$ is more abundant in the cold denser regions, i.e., the main filament, especially when CO has a high depletion factor. This indicates there can be small scale variations in the spatial distribution of $\rm HCO^+$ and $\rm N_2H^+$, even though they are dense gas tracers with similar critical densities. \citet{2015ApJ...799..235G} also suggested that $\rm HCO^+$ better traces intermediate density structures rather than dense structures. We also note from our maps that on large scales, away from the dense filaments, there can be lower density regions that show enhanced $\rm N_2H^+$, e.g., in the northwest region of the map, present over a wide range of CRIRs.


For the maps of the non-colliding case at 5~Myr shown in Figure~\ref{fig:nc-im-co-hcop-n2hp}, we can see similar behaviours as the colliding case. CO depletion can be found along the filament except for the case of $\zeta = 10^{-15}\:{\rm s}^{-1}$. $\rm HCO^+$ and $\rm N_2H^+$ are globally enhanced under stronger CRIR. $\rm HCO^+$ has higher abundances outside the filaments and $\rm N_2H^+$ is enhanced inside the filament. Away from the high density filaments, as in the colliding case, we notice a concentration of $\rm N_2H+$ from a lower density region in the northwest of the map, which is present over the full range of CRIRs.

\subsection{Comparison with Observations}\label{sec:obs}

In Section~\ref{sec:dense-gas}, we have seen that the morphology of certain chemical species approximately follows the position of high density structures. In this section, we further analyse the chemical abundances inside the selected dense clumps. A direct comparison can be made with the observational study of IRDC dense clumps by \citet{Entekhabi2022}. As this observational study selected a common velocity range for all species from the dense gas tracers of ${\rm HCO^+}$(1-0) and ${\rm N_2H^+}$(1-0), we will compare to our results for abundances that use a density threshold of $n_{\rm H} = 10^4\:{\rm cm^{-3}}$ for inclusion. This is 5.5\% of the critical density of ${\rm HCO^+}$(1-0) and 6.6\% of the critical density of ${\rm N_2H^+}$(1-0). However, we will also present another comparison based on synthetic line emission from the full domains without use of a density threshold. One of our main goals here is to examine if the observational data can help constrain the CRIR being experienced by the IRDC.


\begin{figure*}
    \centering
    \includegraphics[width=\linewidth]{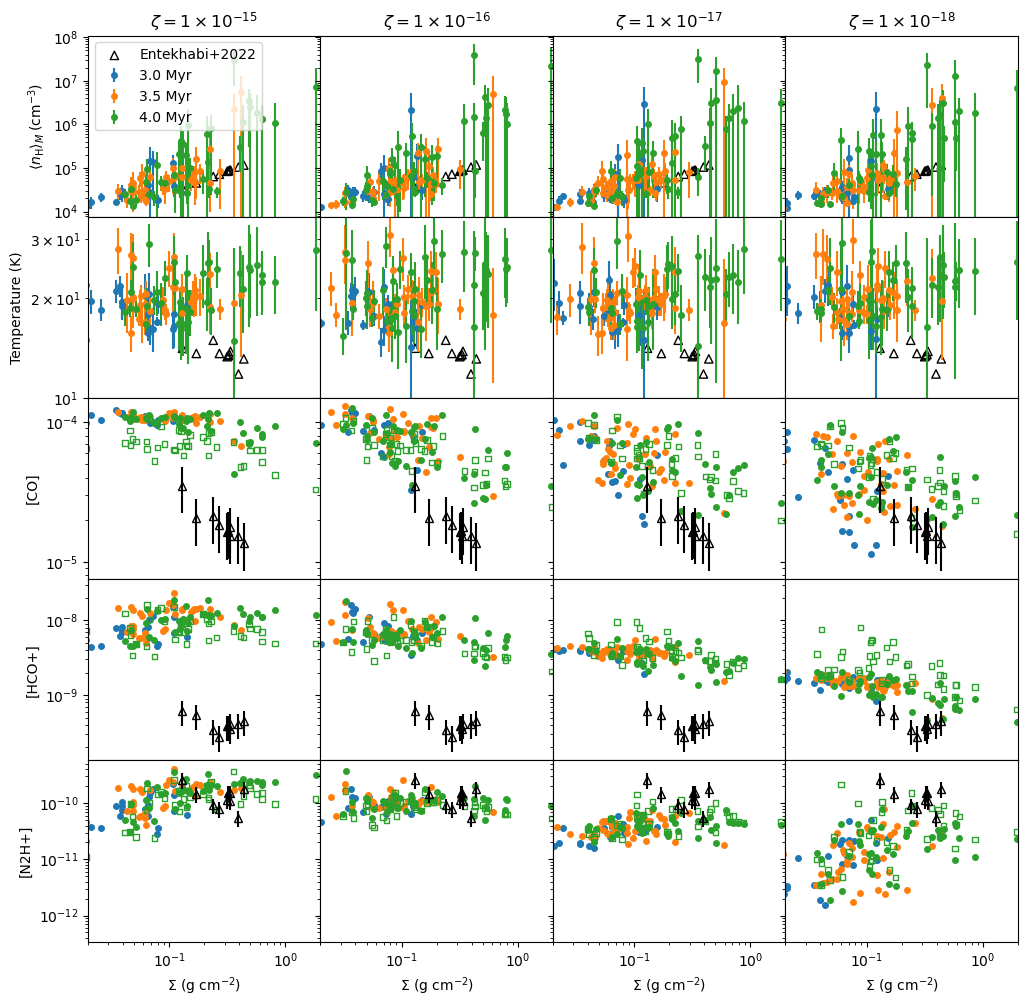}
    \caption{Comparison of simulated GMC-GMC collision produced clumps with observed IRDC clumps. Rows from top to bottom show (1) mass-weighted number density, (2) temperature, (3) CO abundance, (4) $\rm HCO^+$ abundance, and (5) $\rm N_2H^+$ abundance versus mass surface density of selected clumps of the colliding case at 4~Myr. A density threshold of $n_{\rm H} = 10^4\:{\rm cm^{-3}}$ is applied. The columns from left to right show the results under different CRIRs, including $\zeta = 10^{-15}$, $10^{-16}$, $10^{-17}$, and $10^{-18}\:{\rm s^{-1}}$. The blue, orange, and green solid circles show the results at 3, 3.5~Myr, and 4~Myr. The open green squares show abundances estimated from the synthetic line emission maps at 4~Myr. The errorbars plotted in rows (1) and (2) show the standard deviation of the pixels inside each clump. Observational data of IRDC G28.37+0.07 \citep{Entekhabi2022} are plotted in black triangles with errorbars.}
    \label{fig:cc-sc-co-hcop-n2hp}
\end{figure*}

\begin{figure*}
    \centering
    \includegraphics[width=\linewidth]{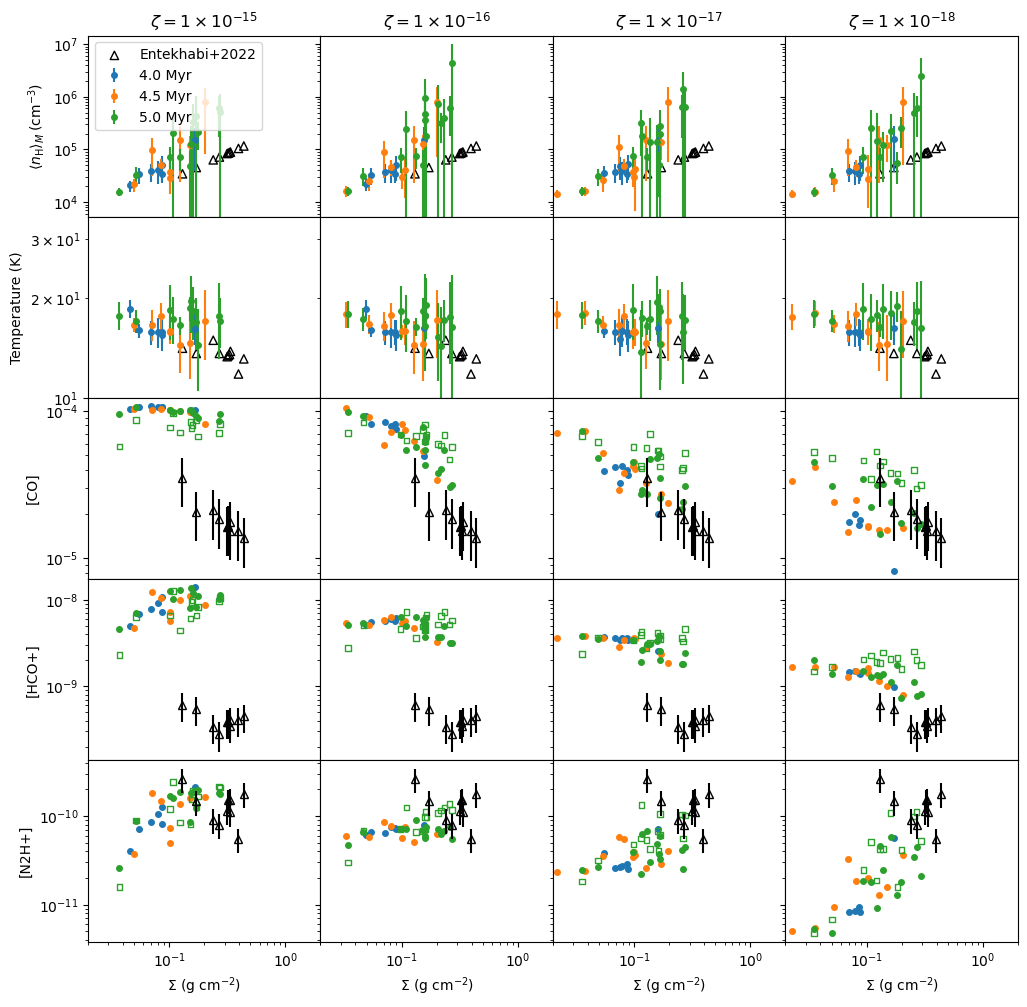}
    \caption{As Figure~\ref{fig:cc-sc-co-hcop-n2hp}, but for the non-colliding cases at 4 (blue), 4.5 (orange) and 5~Myr (green).}
    \label{fig:nc-sc-co-hcop-n2hp}
\end{figure*}

\begin{figure*}
    \centering
    \includegraphics[width=\linewidth]{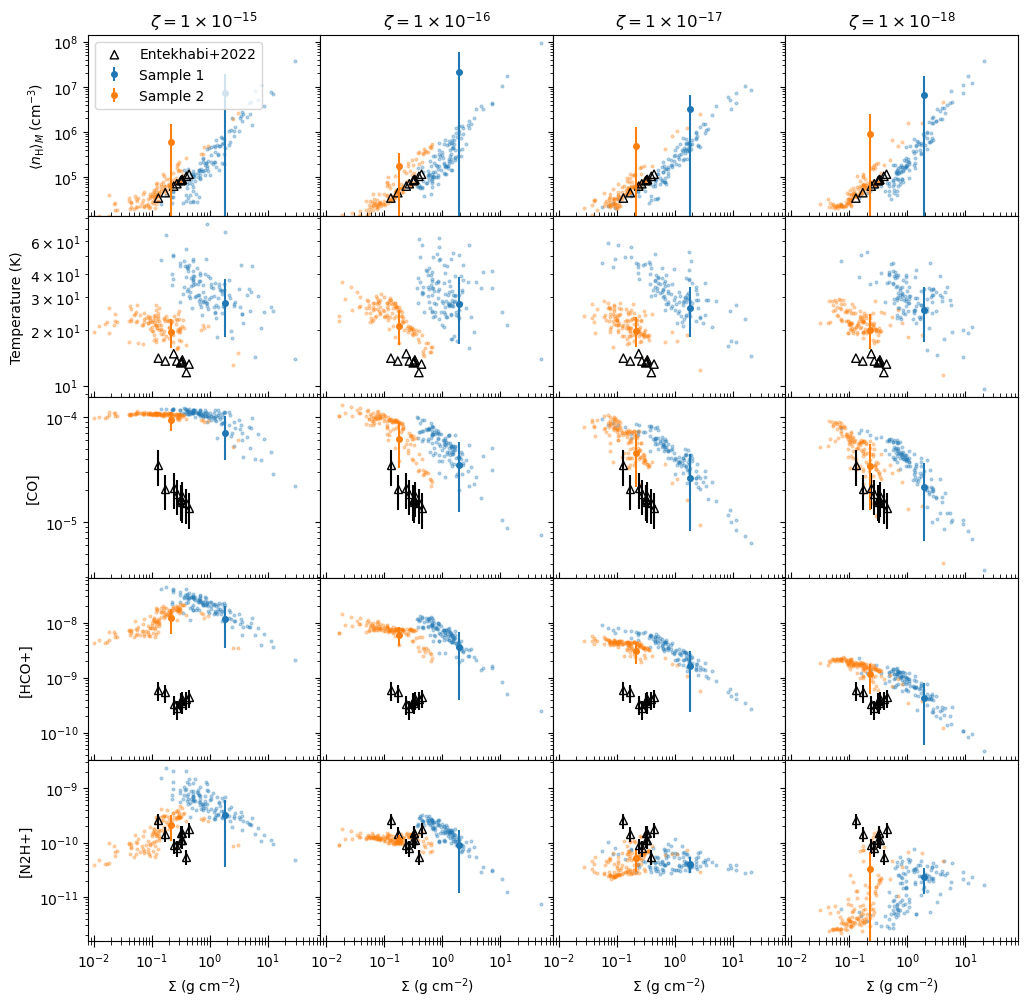}
    \caption{As Figure~\ref{fig:cc-sc-co-hcop-n2hp}, but showing the pixel-wise quantities of two selected sample clumps at 4~Myr. The transparent circles indicate the values in each pixel. The solid circles show the overall average value of each core and the errorbar shows the dispersion of the corresponding transparent circles.}
    \label{fig:cc-sc-pix-co-hcop-n2hp}
\end{figure*}

\begin{figure*}
    \centering
    \includegraphics[width=\linewidth]{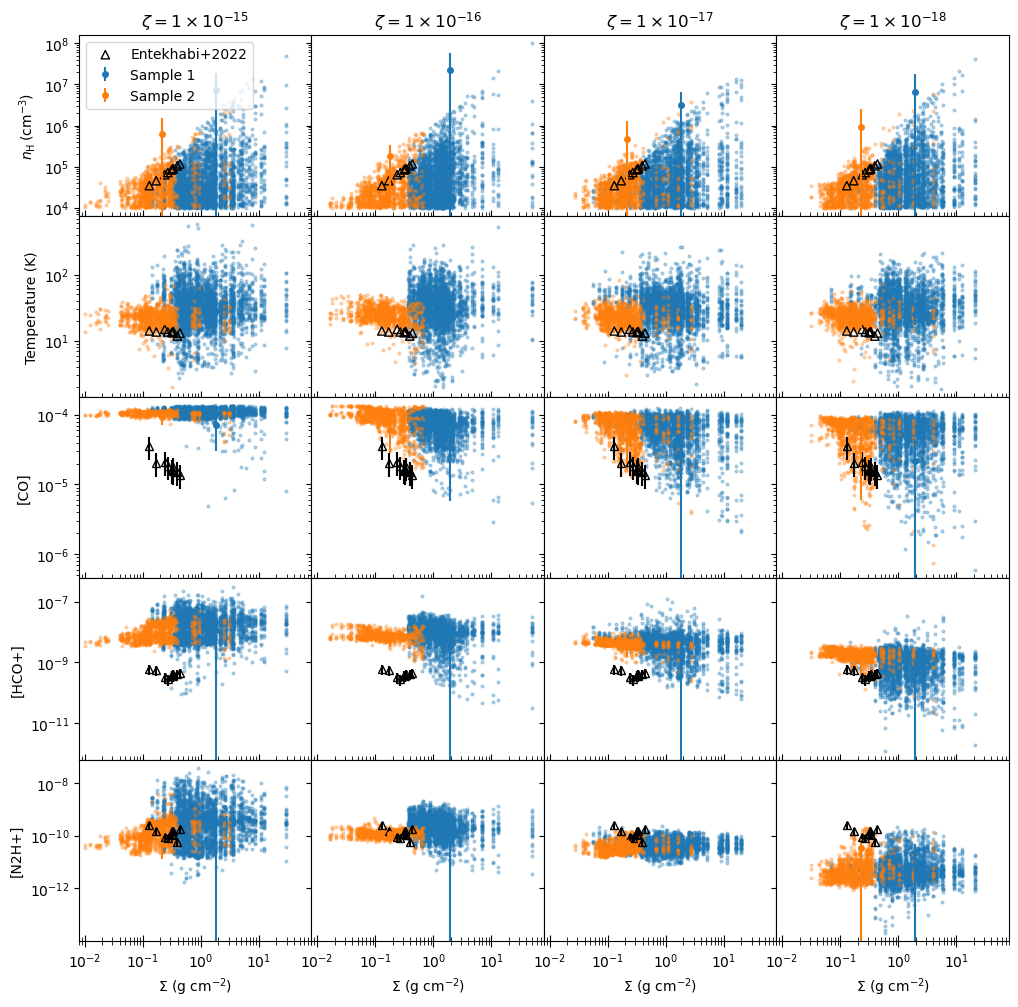}
    \caption{As Figure~\ref{fig:cc-sc-pix-co-hcop-n2hp}, but showing quantities in the 3D cells of two selected sample clumps at 4~Myr. Transparent circles show the values in each cell. Again, the solid circles show the overall average value of each core and the errorbar shows the dispersion of the corresponding transparent circles. Their mass surface densities adopt the ones of the pixels along the line of sight.}
    \label{fig:cc-sc-cell-co-hcop-n2hp}
\end{figure*}

\subsubsection{Comparison to abundances with a density threshold}

Figure~\ref{fig:cc-sc-co-hcop-n2hp} presents a comparison of the properties of the 10 IRDC clumps studied by \citet{Entekhabi2022} with clumps selected from the high density region of the GMC-GMC collision simulations with various CRIRs using a density threshold of $n_{\rm H} = 10^4\:{\rm cm^{-3}}$, as described above. We also show the effects of time evolution in the simulation by plotting results for clumps selected at 3.0, 3.5 and 4.0~Myr. The first two rows of the figure show the number density and temperature of the clumps plotted versus their mass surface densities. We see that the number densities of the simulated clumps overlap reasonably well with the observed sample, whose number density is derived from a simple assumption of spherical geometry. However, the temperatures of our clumps are significantly warmer, i.e., about 20~K, compared to $\lesssim 15\:$K observed in the IRDC clumps. Next, comparing CO abundance, we see that the simulated clumps only achieve the low observed CO abundance (i.e., CO depletion factor of $f_D\sim 3$ to 10) when the CRIR is $\sim 10^{-17}\:{\rm s}^{-1}$ or lower, and even then this only occurs in a minority of the clumps. Considering $\rm HCO^+$, we see that the simulated clumps always have a higher abundance than the observed sample. The closest match is achieved at the lowest CRIR of $10^{-18}\:{\rm s}^{-1}$, but even here the simulated clumps have two or three times higher abundances. Finally, for $\rm N_2H^+$, we see better agreement at the higher values of CRIR, but a relative under abundance in the simulated clumps when lower CRIRs are considered. In general, time evolution from 3 to 4 Myr leads to higher density and slightly warmer clumps, but otherwise does not indicate that the sources will eventually come into better agreement with the data at later times. Thus, we cannot find a consistent solution at a particular value of CRIR for this comparison between simulated and observed clumps. This is potentially due to the fact that the physical property of temperature is systematically higher in these simulated clumps.

Next, in Figure~\ref{fig:nc-sc-co-hcop-n2hp}, we make the same comparison but now with the clumps selected from the non-colliding case, showing time evolution from 4 to 5 Myr. We see that these clumps are colder and so are a closer match to the IRDC clumps. In addition, gas phase CO abundances are in better agreement, if CRIR is $\lesssim 10^{-17}\:{\rm s}^{-1}$. The abundance of $\rm HCO^+$ in the simulated clumps approaches within about a factor of two of the IRDC sample when CRIR is $10^{-18}\:{\rm s}^{-1}$, although is still systematically too high. However, then, as with the colliding case clumps, the simulated ${\rm N_2H^+}$ abundance is too low. However, we notice that some of the most evolved, i.e., at 5~Myr, higher mass surface density clumps are very close, i.e., within a factor of two, in all metrics.

Besides the average abundances of clumps, we also select two sample clumps from the colliding case at 4~Myr and examine the variance of their internal abundances. Figure~\ref{fig:cc-sc-pix-co-hcop-n2hp} shows the mass-weighted number density, temperature and abundances of the pixels in the two clumps. The overall average values of these quantities are also overplotted with their dispersion. For mass-weighted number density, the pixelwise values appear to follow the trend shown in \citet{Entekhabi2022} data even better at high surface density end than the group of clumps shown in Figure~\ref{fig:cc-sc-co-hcop-n2hp}. Although denser pixels usually have lower temperatures, denser clumps are not always cooler. The denser clump sample contains a few intermediate dense pixels with high temperatures that contribute to an overall warmer temperature. From the abundances, we see that they show a similar tendency as Figure~\ref{fig:cc-sc-co-hcop-n2hp}. CO and $\rm HCO^+$ abundances are reduced in high density regions and at lower CRIRs. If we try to match the data of \citet{Entekhabi2022} at the pixel level, these two species still prefer the lowest CRIR of $10^{-18}\:{\rm s}^{-1}$. $\rm N_2H^+$, however, also shows that its abundance decreases as its density increases in the regime with $\Sigma > 0.5\:(cm^{-2})$, which is not clearly shown from the group of clumps. However, the pixelwise comparison still shows that high CRIRs are preferable, although several pixels at the lowest CRIR of $10^{-18}\:{\rm s}^{-1}$ still give a reasonable match.

We also examine the individual 3D cells in two sample cores. Since the cores are selected from the projected plane, all cells along the line of sight of the cores are collected and the corresponding mass surface density in the pixel is treated as the mass surface density of the cells along the line of sight. The density threshold of $n_{\rm H} = 10^4\:{\rm cm^{-3}}$ is still applied and sets a minimum, but we still can see a large number of low density cells contributed by ambient gas no matter how massive the pixels appear. Although the lower density cells expand the distribution of these quantities, their results still look coherent and we still can see the effect of CRIR. Again, CO and $\rm HCO^+$ abundances are reduced when CRIRs are lower. Comparing with \citet{Entekhabi2022} results at the cell level, a preference is still found for the lowest CRIR of $10^{-18}\:{\rm s}^{-1}$, while $\rm N_2H^+$ inclines to higher CRIRs.

\subsubsection{Comparison via synthetic line emission maps}

As described in \S\ref{sec:method}, we carry out radiative transfer of the line emission from the simulations using the code {\tt RADMC-3D}. This is done for $\rm C^{18}O$(1-0), $\rm H^{13}CO^+$(1-0) and $\rm N_2H^+$(1-0), which were the main tracers used in the study of \citet{Entekhabi2022}. The gas densities and temperatures in the high density regions, along with the species abundances (scaled with standard isotopologue ratios), are used as inputs by {\tt RADMC-3D}. Note, that information from all cells is used, i.e., there is no density threshold applied.

\begin{figure*}
    \centering
    \includegraphics[width=\linewidth]{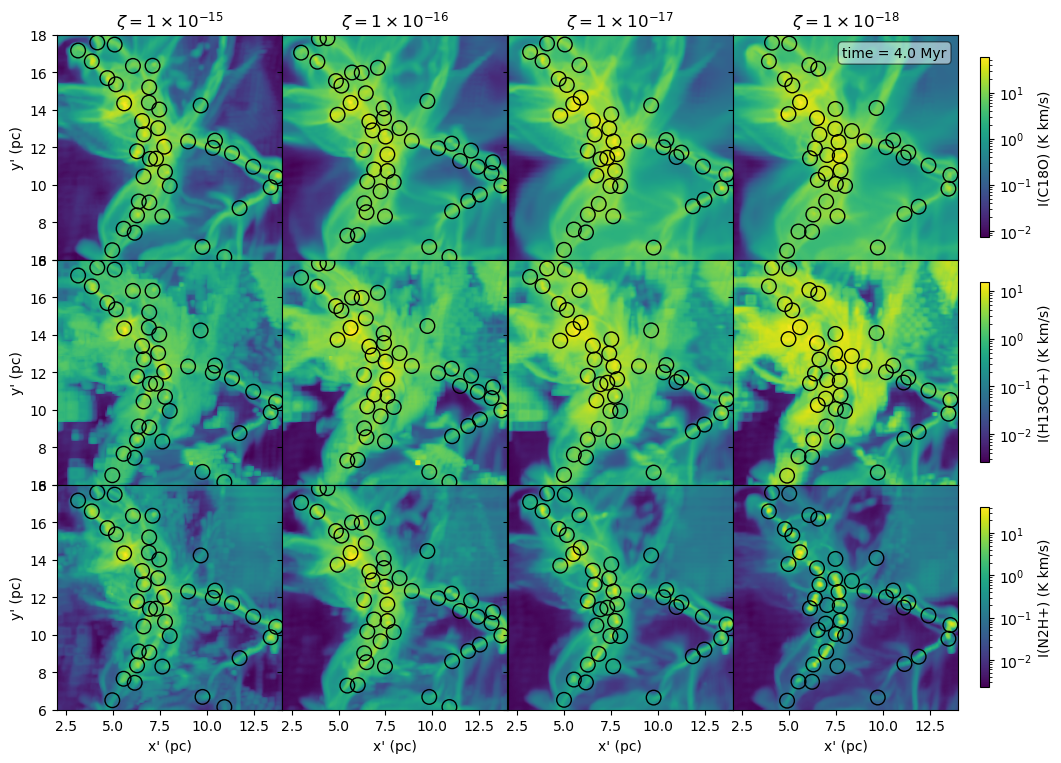}
    \caption{From top to bottom, the integrated intensity maps of C$^{18}$O(1-0), ${\rm H^{13}CO^+}$(1-0) and ${\rm N_2H^+}$(1-0) of the colliding case at 4~Myr for a variety of CRIRs (columns left to right, as labelled). Black circles indicate the clumps selected from the original mass surface density map.}
    \label{fig:cc-im-co-hco-n2h-rt}
\end{figure*}

\begin{figure*}
    \centering
    \includegraphics[width=\linewidth]{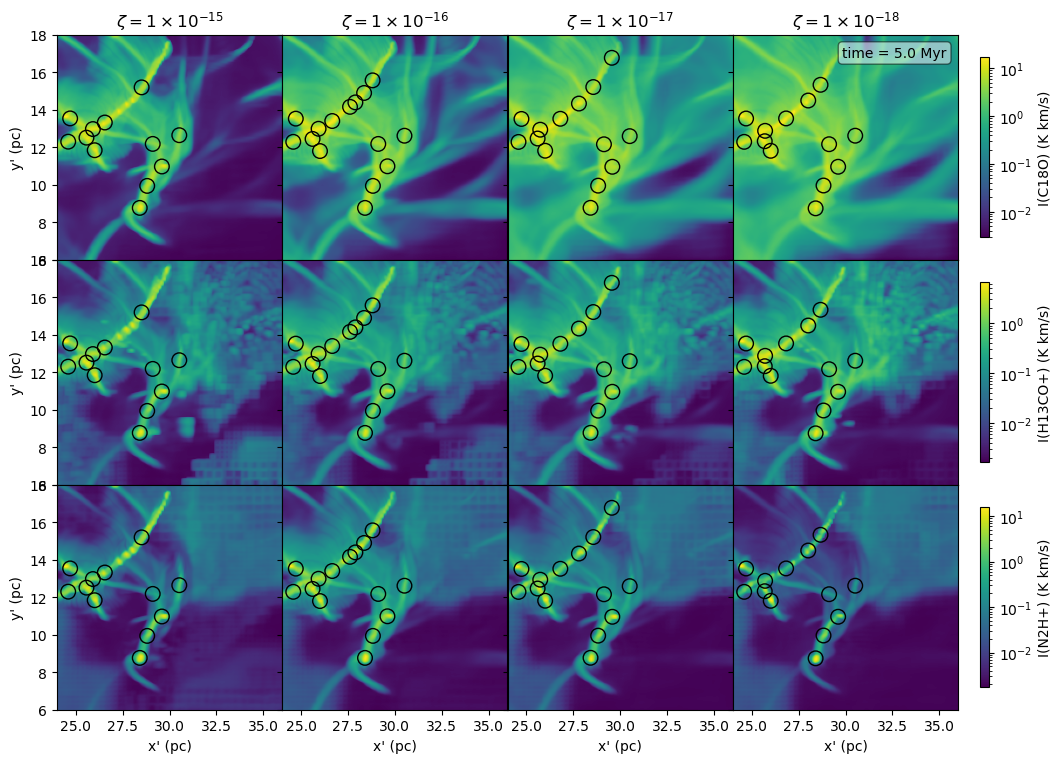}
    \caption{As Figure~\ref{fig:cc-im-co-hco-n2h-rt}, but now for the non-colliding case at 5~Myr.}
    \label{fig:nc-im-co-hco-n2h-rt}
\end{figure*}

\begin{figure*}
    \centering
    \includegraphics[width=\linewidth]{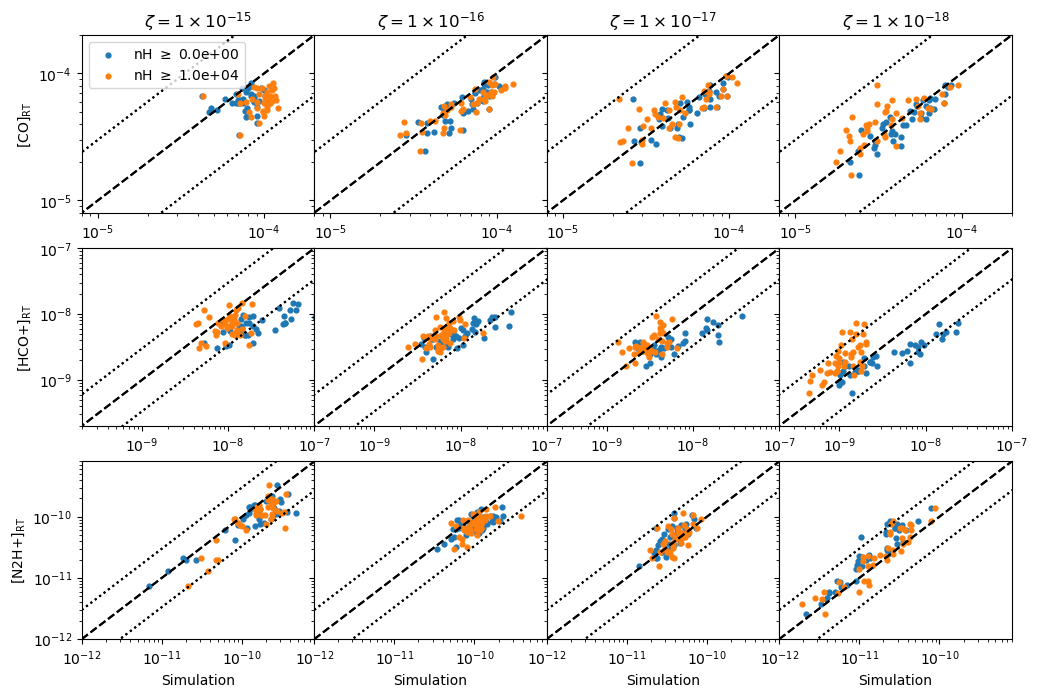}
    \caption{Abundances inferred from synthetic line emission maps for CO, HCO$^+$, and N$_2$H$^+$ versus ``ground truth'' simulations results for clumps selected from the colliding case at 4~Myr, with different CRIR cases in columns, as labelled. The dashed line shows perfect, 1:1 agreement. Dotted lines show factor of three discrepancy to either side. Blue circles are the simulation without any density threshold, while orange cirlces show the results when a density threshold of $n_{\rm H} = 10^4\:{\rm cm^{-3}}$ is applied for inclusion in the calculation. Note the synthetic line emission maps are from transfer through the domain without any density threshold applied.}
    \label{fig:cc-sim-rt}
\end{figure*}

\begin{figure*}
    \centering
    \includegraphics[width=\linewidth]{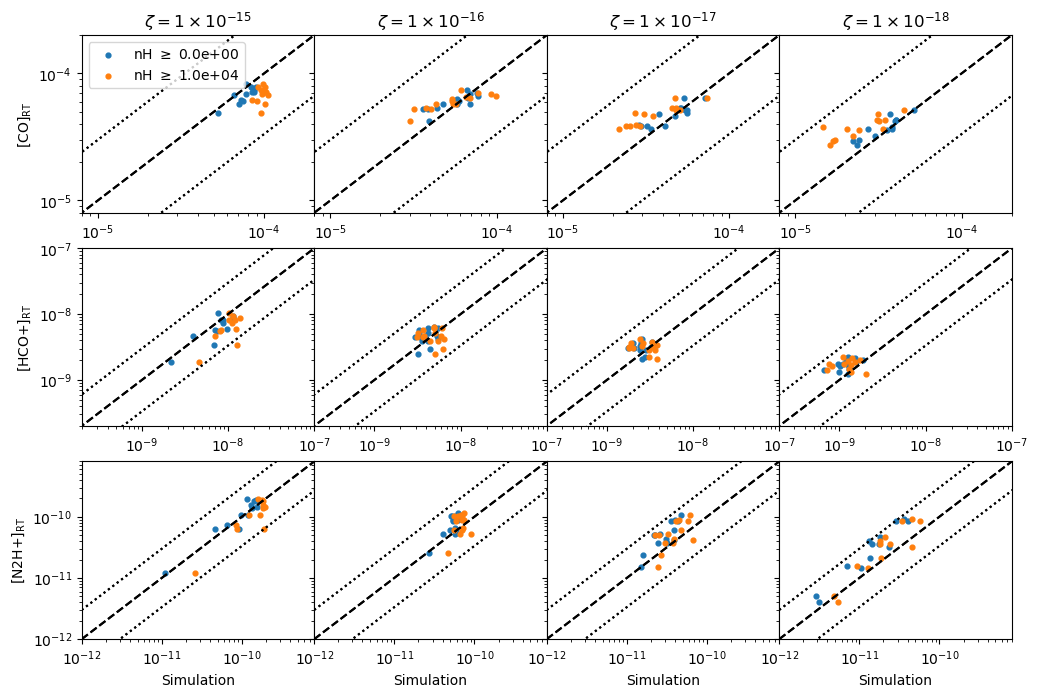}
    \caption{Same as Figure~\ref{fig:cc-sim-rt}, but now for the non-colliding case at 5~Myr.}
    \label{fig:nc-sim-rt}
\end{figure*}

Figures~\ref{fig:cc-im-co-hco-n2h-rt} and \ref{fig:nc-im-co-hco-n2h-rt} show the resulting integrated intensity maps from the colliding and non-colliding simulations, respectively. These data are used to estimate the line emission from the clumps, which are then converted to species abundance using the same assumptions as \citet{Entekhabi2022}, i.e., an assumption of an excitation temperature of 7.5~K. Figures~\ref{fig:cc-sim-rt} and \ref{fig:nc-sim-rt} compare these abundances derived from the line emission and the actual abundances in the simulations, including whether a density threshold is used or not. In general, we see fairly good agreement between these estimates, especially for the non-colliding case. We infer that this is because its clumps have colder, lower density conditions leading to $T_{\rm ex}=7.5\:$K being a better description of the level populations.

The effect of using abundances estimated from the synthetic line emission maps of the final timesteps of the colliding and non-colliding simulations in comparison with the data of \citet{Entekhabi2022} are shown by the open square points in Figures~\ref{fig:cc-sc-co-hcop-n2hp} and \ref{fig:nc-sc-co-hcop-n2hp}, respectively. We see that while there can be modest systematic changes in abundance levels, these do not bring the simulations and the data into significantly closer agreement.





\subsubsection{Comparison to estimates from single grid models}

In observational studies, a typical way to estimate the chemical ages, cosmic-ray ionization rate, and other chemical properties is by comparing the observational data with a series of astrochemical models \citep[e.g.,][]{Entekhabi2022}. However, these single grid models cannot consider the influence of dynamics and it is uncertain how accurate this approach is at inferring chemodynamical history. In this section, we mimic the approach of the observational study by using the density and temperature inferred from the mass surface density map and doing astrochemical modelling with these values. The idea is to understand whether post-processing with single grid models can effectively reproduce the chemical signatures.


\begin{figure*}
    \centering
    \includegraphics[width=\linewidth]{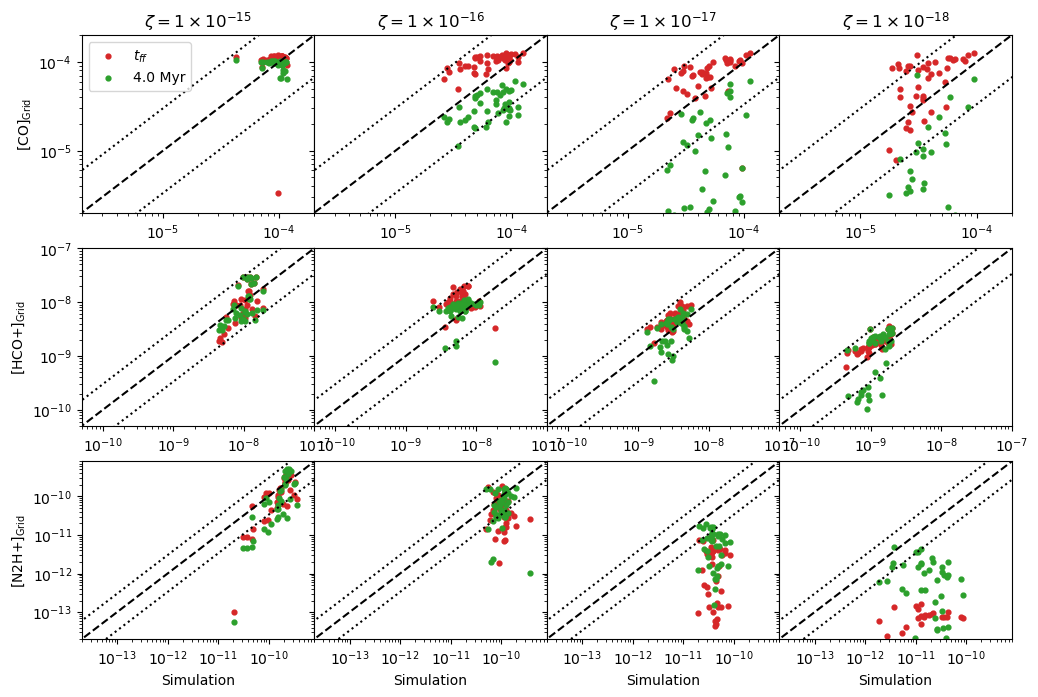}
    \caption{Abundances predicted by single grid models versus the abundances obtained in the clumps in the colliding case at 4~Myr. Note, the models here are based on clump properties when a density threshold of $n_{\rm H} = 10^4\:{\rm cm^{-3}}$ is applied to define the cells for inclusion. The dashed line shows perfect, 1:1 agreement. Dotted lines show a factor of three discrepancy to either side. Red points show the results of the single grid model run for 1 free-fall time. Green points show the results of the single grid model run for 4~Myr, i.e., the same age as the global simulation.
    }
    \label{fig:singlegrid_coll}
\end{figure*}

\begin{figure*}
    \centering
    \includegraphics[width=\linewidth]{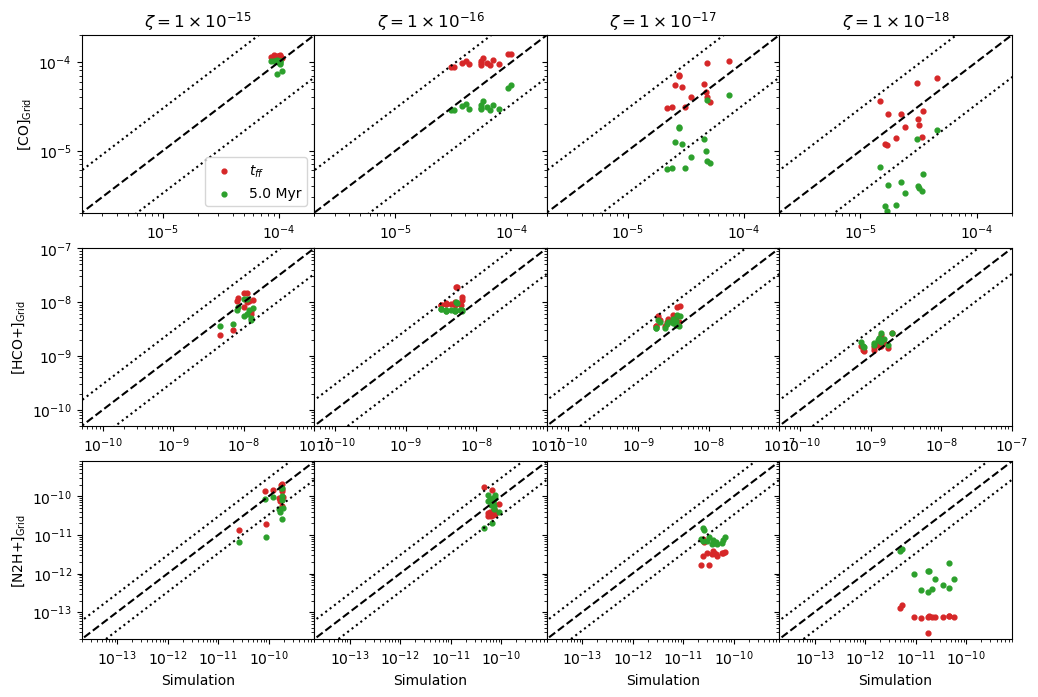}
    \caption{As Figure~\ref{fig:singlegrid_coll}, but now for the non-colliding case at 5~Myr}
    \label{fig:singlegrid_nocoll}
\end{figure*}



We run a single grid astrochemical model using our UCLCHEM reduced network for the density and temperature of each clump and investigate the different CRIR cases. Instead of mass-weighted number density, we use the density estimated by assuming that the mass surface density is from a uniform sphere, which is the assumption made by \citet{Entekhabi2022}. We evaluate the single grid model at two timescales of chemical ages: (1) after one local free-fall time; (2) after the age of the simulation, i.e., 4~Myr for colliding case and 5~Myr for non-colliding case. The abundances of the single grid models and simulations are compared for the colliding case at 4~Myr and the non-colliding case at 5~Myr in Figures~\ref{fig:singlegrid_coll} and \ref{fig:singlegrid_nocoll}, respectively.

For the colliding case, we find that the models evolved for only one local free-fall time can often have quite good agreement in CO abundance between the single grid model and the actual simulation. When there are discrepancies, these tend to be in a direction of a moderate overestimation of CO abundance by the single grid models. This is because the initial condition places all C in gas-phase CO and it takes some time for this to freeze-out onto dust grains. If the single grid model is run for 4~Myr, its CO abundances are generally much lower than seen in the simulation, i.e., a much poorer level of agreement. 
This is because the dense clumps in the simulation have not been at their current dense state for the full 4~Myr of the global evolution of the GMCs, but formed more recently. Only in the highest CRIR case, where CO is kept in the gas phase, is there general agreement, basically because the CO abundance is not changing much in time and stays at a level near $10^{-4}$. Thus if a single grid model is to be used, then a timescale closer to 1 or a few free-fall times is preferred to explain the properties of these particular simulated clumps. 
For $\rm HCO^+$, we see a similar behaviour, i.e., the single grid results for one free-fall time evolution are a reasonable predictor of the actual abundances. For $\rm N_2H^+$ the situation is different. For CRIRs of $10^{-16}\:{\rm s}^{-1}$ and lower, one free-fall time is not enough to build up the abundance of $\rm N_2H^+$ seen in the simulation (recall the initial condition is for all N to be in atomic form). 
Even a 4~Myr time evolution is insufficient when CRIR is $10^{-17}\:{\rm s}^{-1}$ and lower. This indicates that the actual dynamical history of the clumps, i.e., evolving from a warmer, lower density state, was important in setting elevated abundances of $\rm N_2H^+$ and this evolution is not captured in a single grid model that maintains fixed density and temperature based on current conditions. Examining the results shown in Figure~\ref{fig:singlegrid_nocoll}, we see that the above trends also hold for the clumps in the non-colliding case.

\section{Conclusions}
\label{sec:conclusion}

We have carried out a series of simulations with active chemistry using a gas-grain network that includes cosmic ray induced desorption processes to study the chemical evolution inside colliding and non-colliding giant molecular clouds (GMCs). We have investigated the influence of cosmic ray ionization rate by testing four different values from $10^{-18}$ to $10^{-15}\:{\rm s^{-1}}$. We note the same physical model has been run, i.e., with heating and cooling rates approximated to be independent of CRIR, which helps isolate the effects of its variation on the chemical evolution.



On large scales, we have examined several species to study the dominant reservoirs of elemental C and O in the diffuse gas and the molecular clouds. We follow the transition of carbon from C$^+$ to atomic C to CO as material becomes denser and more well shielded from the back FUV radiation field, noting the growing importance and thickness of the atomic C layer with increasing CRIR. For oxygen, in addition to CO, atomic O and $\rm H_2O$ ice are the most important species in and around the GMCs.


We have also examined the chemical abundances in high density regions in the GMCs. The colliding GMCs produce greater concentrations of dense and moderately warmer ($\sim 20\:$K) gas, than the non-colliding GMCs.
We first examined the freeze-out of CO onto dust grains, i.e., via the CO depletion factor, which occurs in cold, dense conditions. Especially for cases with lower CRIRs, most CO freezes out leaving only $\sim 10$ to 20\% remaining in the gas phase. The dependencies of CO depletion factor with mass surface density and temperature are sensitive the CRIR, thus providing potential methods for its measurement.

We then considered abundances of dense gas tracers $\rm HCO^+$ and $\rm N_2H^+$, in addition to CO, in samples of dense clumps selected via the dendrogram algorithm from mass surface density maps. We found that the abundances of these molecular ions vary systematically with CRIR, again providing another potential metric for its estimation. However, from a comparison to the observed abundances of these species in 10 IRDC clumps by \citet{Entekhabi2022}, it is difficult to find full self-consistent solutions that match all three species abundances, except in some relatively cold ($\sim15\:$K) and older clumps that are present in the non-colliding GMCs and if the CRIR is relatively low, i.e., $\sim 10^{-18}\:{\rm s}^{-1}$. This is consistent with the fact that the observed clumps also have relatively cold temperatures. A comparison to abundances estimated via synthetic line emission maps (and assumption of excitation temperatures of 7.5~K) does not change the above conclusions. An investigation in the validity of a single grid chemical model reveals the limitations of this method. If it is to be used, then, for the clumps in our simulated clouds, then a timescale of one to a few free-fall times of present density clumps appears most valid. However, we note that our simulations used a relatively weak ($10\:{\rm \mu G}$) initial magnetic field, so such fast collapsing clumps may not be generally valid.

There are a number of caveats with the presented models. As mentioned, they are for a given physical set-up, i.e., with relatively weak magnetic fields. Localized star formation feedback is not included, so the simulations are applicable only to early stage, pre-stellar clouds, clumps and cores, e.g., conditions found in IRDCs. As also mentioned, we have adopted a fixed set of heating and cooling rates that are independent of CRIR. While this choice means we have a fixed physical evolution with which to explore variations of astrochemical evolution, it does mean that our temperatures are overestimated in the dense gas of the low CRIR cases, i.e., the $10^{-17}$ and $10^{-18}\:{\rm s}^{-1}$ cases. This may affect the detailed comparison of the abundance properties of these dense clumps with the observational results of \citet{Entekhabi2022}. A future study will incorporate more general heating and cooling functions and also explore effects of local variation of CRIRs, e.g., due to shielding and attenuation, which may affect the densest structures. Another limitation is that grain sputtering processes that will be enhanced in shocked regions have also not yet been implemented in our modeling. Finally, it is acknowledged that there are significant uncertainties in aspects of astrochemical modeling, including the processes by which cosmic rays interact with the ice species to induce reactions and desorption. However, it is hoped that by studies such as the one presented here, including detailed comparison to observed systems, such aspects may be better understood and refined in future models.

\section*{Acknowledgements}

The simulations were performed on resources provided by the National Academic Infrastructure for Supercomputing in Sweden (NAISS) and the Swedish National Infrastructure for Computing (SNIC) at C3SE/Vera and NSC/Tetralith partially funded by the Swedish Research Council through grant agreements no. 2022-06725 and no. 2018-05973. JCT acknowledges support from VR grant 2017-04522 (Eld ur is) and ERC Advanced Grant 788829 (MSTAR). SV and JH acknowledge support from ERC Grant 833460 (MOPPEX). SV and JCT also acknowledge support from a Royal Society International Exchange Grant. We thank Paola Caselli, Giuliana Cosentino, Negar Entekhabi, Rub\'en Fedriani, Brandt Gaches, Prasanta Gorai, Chi-Yan Law and Catherine Walsh for helpful discussions. This research made use of astrodendro, a Python package to compute dendrograms of Astronomical data (http://www.dendrograms.org/) and yt \citep[https://yt-project.org/,][]{Turk2011} to analysis Enzo ouput. 

\section*{Data Availability}

The data underlying this article will be shared on reasonable request to the corresponding author.



\bibliographystyle{mnras}
\bibliography{example} 




\appendix

\section{Maps of additional species}

\begin{figure*}
    \centering
    \includegraphics[width=\linewidth]{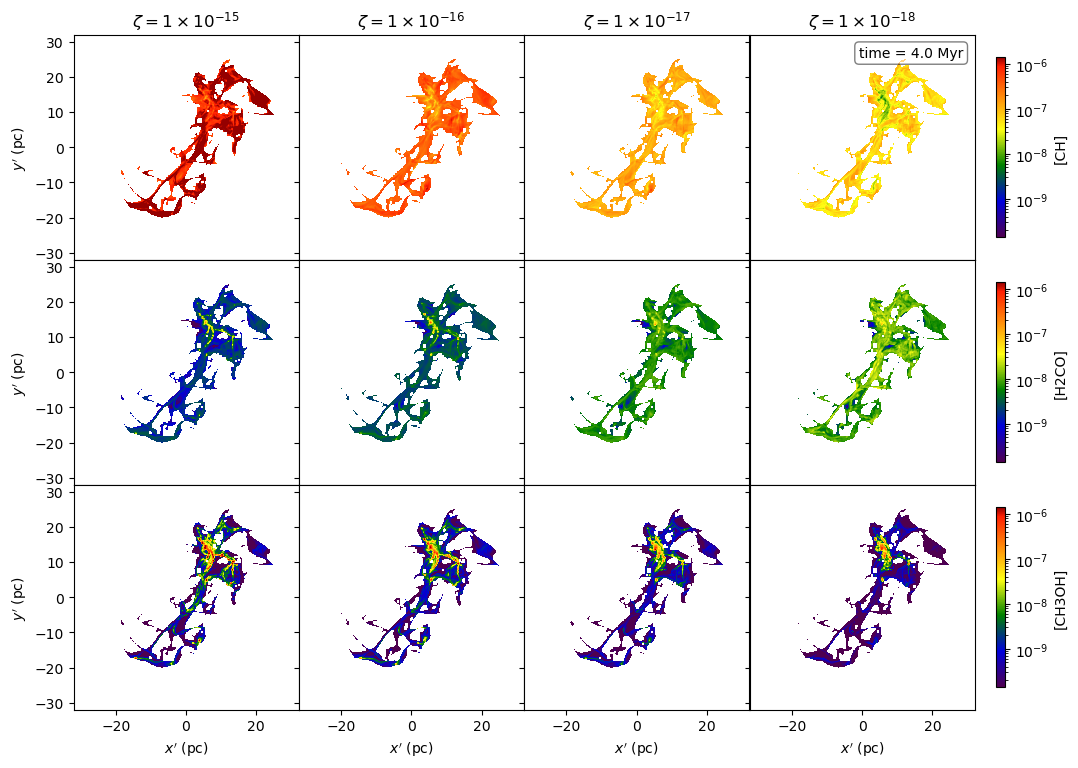}
    \caption{The rows from top to bottom show the maps of CH, H$_2$CO, CH$_3$OH of the colliding case with a density threshold of $n_{\rm H} = 10^3\:{\rm cm}^{-3}$ at 4~Myr. The columns from left to right show the results under different CRIRs ($\zeta = 10^{-15}$, $10^{-16}$, $10^{-17}$ and $10^{-18}\:{\rm s^{-1}}$).}
    \label{fig:cc-im-ch-h2co-ch3oh}
\end{figure*}

\begin{figure*}
    \centering
    \includegraphics[width=\linewidth]{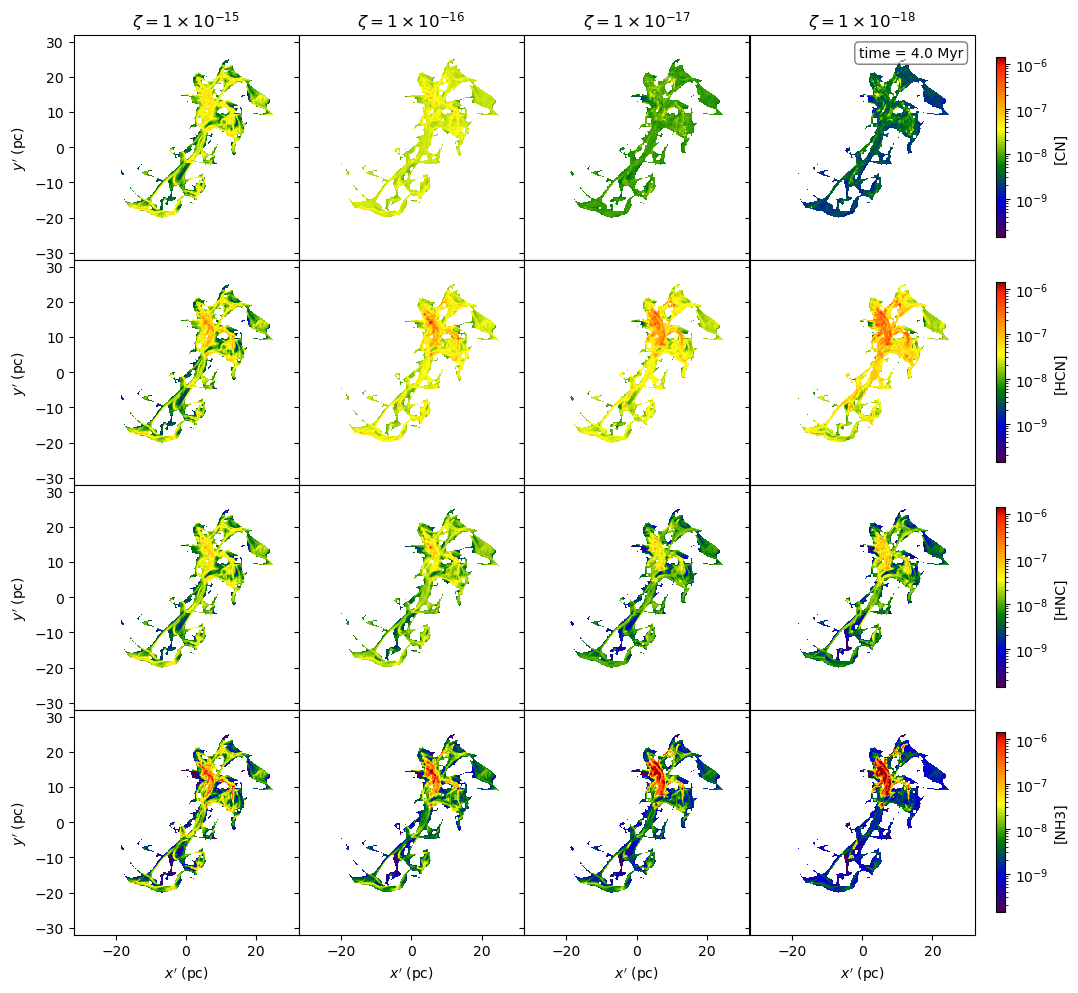}
    \caption{The rows from top to bottom show the maps of CN, HCN, HNC, NH$_3$ of the colliding case with a density threshold of $n_{\rm H} = 10^3\:{\rm cm}^{-3}$ at 4~Myr. The columns from left to right show the results under different CRIRs ($\zeta = 10^{-15}$, $10^{-16}$, $10^{-17}$ and $10^{-18}\:{\rm s^{-1}}$).}
    \label{fig:cc-im-cn-hcn-hnc-nh3}
\end{figure*}

Here, for completeness, in this Appendix we present global region maps of the 4~Myr GMC-GMC collision simulation of the following species: CH, H$_2$CO, CH$_3$OH (Figure~\ref{fig:cc-im-ch-h2co-ch3oh}); CN, HCN, HNC, NH$_3$ (Figure~\ref{fig:cc-im-cn-hcn-hnc-nh3}).

\bsp	
\label{lastpage}
\end{document}